\documentclass[onecolumn]{IEEEtran}
\usepackage{amsmath,amssymb,bm,cite,color,amsthm}
\usepackage[thicklines]{cancel}
\usepackage{enumerate}
\usepackage[linesnumbered,ruled]{algorithm2e}
\usepackage{hyperref}
\usepackage{cleveref}

\newtheorem{theorem}{Theorem}[section]
\newtheorem{lemma}{Lemma}[section]

\newtheorem{clmr}{Claim}
\newtheorem*{clm}{Claim}

\newtheorem{proposition}{Proposition}[section]

\newtheorem{definition}{Definition}[section]

\newtheorem{observation}{Observation}[section]
\newtheorem{example}{Example}[section]
\newtheorem{remark}{Remark}[section]

\newtheorem{question}{Question}

\newcommand\nc\newcommand
\nc{\cA}{\mathcal{A}}\nc{\cB}{\mathcal{B}}\nc{\cC}{\mathcal{C}}\nc{\cD}{\mathcal{D}}
\nc{\cE}{\mathcal{E}}\nc{\cF}{\mathcal{F}}\nc{\cG}{\mathcal{G}}\nc{\cH}{\mathcal{H}}
\nc{\cI}{\mathcal{I}}\nc{\cJ}{\mathcal{J}}\nc{\cK}{\mathcal{K}}\nc{\cL}{\mathcal{L}}
\nc{\cM}{\mathcal{M}}\nc{\cN}{\mathcal{N}}\nc{\cO}{\mathcal{O}}\nc{\cP}{\mathcal{P}}
\nc{\cQ}{\mathcal{Q}}\nc{\cR}{\mathcal{R}}\nc{\cS}{\mathcal{S}}\nc{\cT}{\mathcal{T}}
\nc{\cU}{\mathcal{U}}\nc{\cV}{\mathcal{V}}\nc{\cW}{\mathcal{W}}\nc{\cX}{\mathcal{X}}
\nc{\cY}{\mathcal{Y}}\nc{\cZ}{\mathcal{Z}}

\nc{\bba}{\mathbf{a}}\nc{\bbb}{\mathbf{b}}\nc{\bbc}{\mathbf{c}}\nc{\bbd}{\mathbf{d}}
\nc{\bbe}{\mathbf{e}}\nc{\bbf}{\mathbf{f}}\nc{\bbg}{\mathbf{g}}\nc{\bbh}{\mathbf{h}}
\nc{\bbi}{\mathbf{i}}\nc{\bbj}{\mathbf{j}}\nc{\bbk}{\mathbf{k}}\nc{\bbl}{\mathbf{l}}
\nc{\bbm}{\mathbf{m}}\nc{\bbn}{\mathbf{n}}\nc{\bbo}{\mathbf{o}}\nc{\bbp}{\mathbf{p}}
\nc{\bbq}{\mathbf{q}}\nc{\bbr}{\mathbf{r}}\nc{\bbs}{\mathbf{s}}\nc{\bbt}{\mathbf{t}}
\nc{\bbu}{\mathbf{u}}\nc{\bbv}{\mathbf{v}}\nc{\bbw}{\mathbf{w}}\nc{\bfx}{\mathbf{x}}
\nc{\bby}{\mathbf{y}}\nc{\bbz}{\mathbf{z}}

\nc{\bbA}{\mathbf{A}}\nc{\bbB}{\mathbf{B}}\nc{\bbC}{\mathbf{C}}\nc{\bbD}{\mathbf{D}}
\nc{\bbE}{\mathbf{E}}\nc{\bbF}{\mathbf{F}}\nc{\bbG}{\mathbf{G}}\nc{\bbH}{\mathbf{H}}
\nc{\bbI}{\mathbf{I}}\nc{\bbJ}{\mathbf{J}}\nc{\bbK}{\mathbf{K}}\nc{\bbL}{\mathbf{L}}
\nc{\bbM}{\mathbf{M}}\nc{\bbN}{\mathbf{N}}\nc{\bbO}{\mathbf{O}}\nc{\bbP}{\mathbf{P}}
\nc{\bbQ}{\mathbf{Q}}\nc{\bbR}{\mathbf{R}}\nc{\bbS}{\mathbf{S}}\nc{\bbT}{\mathbf{T}}
\nc{\bbU}{\mathbf{U}}\nc{\bbV}{\mathbf{V}}\nc{\bbW}{\mathbf{W}}\nc{\bfX}{\mathbf{X}}
\nc{\bbY}{\mathbf{Y}}\nc{\bbZ}{\mathbf{Z}}

\nc{\sA}{\mathsf{A}}\nc{\sB}{\mathsf{B}}\nc{\sC}{\mathsf{C}}\nc{\sD}{\mathsf{D}}
\nc{\sE}{\mathsf{E}}\nc{\sF}{\mathsf{F}}\nc{\sG}{\mathsf{G}}\nc{\sH}{\mathsf{H}}
\nc{\sI}{\mathsf{I}}\nc{\sJ}{\mathsf{J}}\nc{\sK}{\mathsf{K}}\nc{\sL}{\mathsf{L}}
\nc{\sM}{\mathsf{M}}\nc{\sN}{\mathsf{N}}\nc{\sO}{\mathsf{O}}\nc{\sP}{\mathsf{P}}
\nc{\sQ}{\mathsf{Q}}\nc{\sR}{\mathsf{R}}\nc{\sS}{\mathsf{S}}\nc{\sT}{\mathsf{T}}
\nc{\sU}{\mathsf{U}}\nc{\sV}{\mathsf{V}}\nc{\sW}{\mathsf{W}}\nc{\sX}{\mathsf{X}}
\nc{\sY}{\mathsf{Y}}\nc{\sZ}{\mathsf{Z}}

\newcommand{\mathset}[1]{\left\{#1\right\}}

\newcommand{\abs}[1]{\left|#1\right|}
\newcommand{\ceilenv}[1]{\left\lceil #1 \right\rceil}
\newcommand{\floorenv}[1]{\left\lfloor #1 \right\rfloor}
\newcommand{\parenv}[1]{\left( #1 \right)}
\newcommand{\sparenv}[1]{\left[ #1 \right]}

\nc{\set}[1]{\llbracket #1 \rrbracket}

\newcommand{\Sum}[1]{\mathsf{Sum}\left(#1\right)}
\newcommand{\vt}{\mathsf{VT}}
\newcommand{\dvt}{\mathsf{DVT}}

\newcommand{\e}{\mathsf{e}}

\title{Correcting Bursty/Localized Deletions: A New Error-Position-Estimation Code}
\author{Zuo~Ye, Yubo Sun and Gennian Ge%
\thanks{The research of Y. Sun and G. Ge was supported by the National Key Research and Development Program of China under Grant 2020YFA0712100, the National
Natural Science Foundation of China under Grant 12231014, and Beijing Scholars Program.

Z. Ye is with the Institute of Mathematics and Interdisciplinary Sciences, Xidian University, 
Xian 710126, China. Email: yezuo@xidian.edu.cn.

Y. Sun and G. Ge are with the School of Mathematical Sciences, Capital Normal University, Beijing 100048, China, Emails: 2200502135@cnu.edu.cn, gnge@zju.edu.cn.
}
}

\begin{document}
\maketitle

\begin{abstract}
Codes correcting bursts of deletions and localized deletions have garnered significant research interest in recent years. One of the primary objectives is to construct codes with minimal redundancy. Currently, the best known constructions of $q$-ary codes correcting a burst of at most $t$ deletions ($(\le t)$-burst-deletion correcting codes) achieve redundancy $\log n+8\log\log n+o(\log\log n)$ (for any $q$ and $t$) or $\log n+t\log\log n+O(1)$ (for even $q$). For codes correcting single $t$-localized-deletion ($t$-localized-deletion correcting codes), state-of-the-art constructions attain redundancy $\log n+O\parenv{t(\log\log n)^2}$ (for any $q$ and $t$) or $\log n+2t\log\log n+O(1)$ (for even $q$). Here, $n$ denotes the code-length, and $q$ and $t$ are fixed. These codes employ a position-estimation component to approximate error positions, augmented by additional constraints that enable error-correction given the information about error positions.

In this work, we select codewords from the set of sequences whose differential sequences are strong-$(\ell,\epsilon)$-locally-balanced. By imposing a VT-type constraint and an $L_1$-weight constraint on the differential sequences of codewords, we construct novel position-estimation codes. When $q\ge 2$ and $t<q$, or $q$ is even and $t<2q$, this approach gives a $q$-ary $(\le t)$-burst-deletion correcting code and a $t$-localized-deletion correcting code with redundancy $\log n+(t-1)\log\log n+O(1)$. In addition to improving previous redundancy, the method is new and our position-estimation codes are simpler than those in previous works. Finally, we give an efficient encoder to encode an arbitrary input sequence into a sequence whose differential sequence is strong-$(\ell,\epsilon)$-locally-balanced. To our knowledge, no prior algorithm for this specific task has been reported.
\end{abstract}


\section{Introduction}\label{sec_introduction}
%%%%%%%%%%%%%%%%%%%%%%%%%%%%%%%%%%%%%%%%%%%%%%%%%
A deletion in a sequence refers to the removal of a symbol within that sequence. Codes correcting deletions were first explored in the 1960s \cite{VT1965,VL1966}. It was proved in \cite{VL1966} that the binary code proposed in \cite{VT1965} (VT code) can correct a single deletion. This construction was later generalized to non-binary alphabets in 1984 \cite{Tenengolts1984}. Both codes in \cite{VL1966} and \cite{Tenengolts1984} achieve asymptotically optimal redundancy. Driven by applications in various communication and storage systems (e.g, DNA-based storage \cite{Yazdi2015IEEE}, racetrack memory \cite{Sima2023IT} and document exchange \cite{Haeupler2019FOCS}), deletion correcting codes have been extensively investigated in the past decade. There are three main lines of works on constructing codes combating adversarial deletions. The first one addresses the case where the positions of deleted symbols are mutually independent. As this falls outside the scope of this paper, we omit detailed discussion. Interested readers may refer to \cite{Sima2020isit_deletion,Bruck2020isit,JinRyan2020isit,Sima2021it,Guruswami2021it,Song2022IT,Song2023IT} and references therein.

The second line focuses on codes correcting single burst-deletion, i.e. deletions occurring in consecutive positions. Levenshtein’s work in 1967 \cite{VL1967} pioneered this area, constructing a binary length-$n$ code capable of correcting at most two adjacent deletions and the redundancy is $\log n+1$. For general $t\ge 2$, Schoeny \textit{et al} in 2017 \cite{Schoeny2017it} constructed a binary $t$-burst-deletion (a burst of \emph{exactly} $t$ deletions) correcting code with redundancy $\log n+(t-1)\log\log n+O(1)$\footnote{Throughout this paper, the alphabet size and the total number of deletions are assumed to be fixed constants.}, which improves the results in \cite{LingChengisit}. Subsequent two works \cite{Schoeny2017,Saeki2018ISITA} generalized this construction to non-binary alphabets. The redundancy of $t$-burst-deletion correcting codes was proved to be lower bounded by $\log n+\Omega(1)$ \cite{Schoeny2017it,Saeki2018ISITA}. The asymptotic tightness of this lower bound was confirmed by a recent work of Sun \textit{et al} \cite{Yubo2025IT}, where a code with redundancy $\log n+O(1)$ was constructed.

In \cite{Schoeny2017it}, Schoeny \textit{et al} also studied binary $(\le t)$-burst-deletion correcting codes, that is, codes that can correct a burst of \emph{at most} $t$ deletions. The redundancy of their constructed code is $(t-1)\log n+\parenv{t(t-1)/2-1}\log\log n+O(1)$. This construction was later improved and extended to general alphabets by several researchers \cite{Ryan2018IT,Lenz2020ISIT,JinRyan2020isit,Song2023IT,Shuche2024IT,Song2025Entropy,Yubo2025IT}. For any $q\ge 2$, the best known redundancy $\log n+8\log\log n+o(\log\log n)$ was attained by constructions given in \cite{Song2023IT,Song2025Entropy}. When $q\ge 2$ is even and $2<t\le 8$, the construction in \cite[Section VII-C]{Yubo2025IT} achieves the best known redundancy $\log n+t\log\log n+O(1)$. It was proved in \cite{Shuche2024IT} that the redundancy of a $(\le t)$-burst-deletion correcting code is bounded from below by $\log n+\Omega(1)$. When $t=2$, the code in \cite{VL1967} shows that this lower bound is tight up to a constant for the binary alphabet. Currently, it is not known if this bound is asymptotically tight for general $t$ and general alphabets. For a summary of results on $t$-burst-deletion and $(\le t)$-burst-deletion correcting codes, we refer to \cite[Table I]{yezuo202409}.

The third line of works focuses on the $t$-localized-deletion error, which is a generalization of the burst-deletion error model. Briefly speaking, a $t$-localized-deletion in a sequence refers to deleting at most $t$ symbols in a length-$t$ window of the sequence and the position of the window is not known a priori. The study of this kind of error was initiated by Schoeny \textit{et al} in \cite{Schoeny2017it}. For $t=3$ and $t=4$, they designed binary codes with $4\log n+2\log\log n+6$ and $7\log n+2\log\log n+4$ redundant bits, respectively. Following this work, Bitar \textit{et al} \cite{RawadBitar2021isit} further investigated constructions of binary codes correcting a $t$-localized-deletion and constructed a code with $\log n+O\parenv{t\log^2(t\log n)}$ redundant bits under the assumption that $t=O\parenv{n/\log^2 n}$. For fixed $t$ (the regime of interest in this paper), the redundancy reduces to $\log n+O\parenv{t(\log\log n)^2}$. Later, Sun \textit{et al} \cite[Section VII-D]{Yubo2025IT} constructed a $q$-ary $t$-localized-deletion correcting code with redundancy $\log n+2t\log\log n+O(1)$ for any fixed even $q$. Since a $t$-localized-deletion correcting code is necessarily a $(\le t)$-burst-deletion correcting code, the redundancy of a $t$-localized-deletion correcting code inherits the lower bound $\log n+\Omega(1)$. In addition to codes correcting single $t$-localized-deletions mentioned above, Sima \textit{et al} also constructed a code which can correct multiple $t$-localized-deletions with low redundancy \cite{JinRyan2020isit}.

Except for constructions in \cite{JinRyan2020isit}, \cite{VL1967}, \cite{LingChengisit}  and \cite[Section VI-C]{Yubo2025IT},  all codes listed above share a common spirit: the error-correcting algorithms first estimate the approximate error positions and then correcting errors with the given information about error positions. According to their position-estimation techniques, these works can be categorized into two classes. The idea in \cite{Schoeny2017it}, \cite{Ryan2018IT} and \cite[Section IV]{Shuche2024IT} is to represent a codeword as an array. A VT-type constraint and a run-length-limited constraint (or certain ``balance" constraint or pattern-limited constraint) are imposed on the first row. Then one can approximately locate deletions in each row by correcting deletions in the first row. As for constructions in \cite{Schoeny2017it} and \cite{Ryan2018IT}, one has to design a position-estimation code for each possible burst-length $t^\prime\le t$. This increases the overall redundancy. To overcome this drawback, Lenz and Polyanskii \cite{Lenz2020ISIT} developed a novel position-estimation method to construct binary $(\le t)$-burst-deletion correcting codes. They selected codewords from the set of $(\bbp,\delta)$-dense sequences and associated a sequence $a_{\bbp}(\bfx)$ to each codeword $\bfx$. Then they imposed a VT-type constraint on $a_{\bbp}(\bfx)$ and a constraint on the number of sequences $\bbp$ in the codeword $\bfx$. This gives a single position-estimation code to handle all possible burst-lengths, which greatly decreases the overall redundancy. Later, \cite{Song2023IT}, \cite[Section V]{Shuche2024IT} and \cite[Section VII-C]{Yubo2025IT} gave constructions of $(\le t)$-burst-deletion correcting codes based on the position-estimation method developed in \cite{Lenz2020ISIT}. Construction in \cite{Song2025Entropy} generalizes the method in \cite{Lenz2020ISIT} to general alphabets. The work \cite{RawadBitar2021isit} (and \cite{Yubo2025IT}) generalizes the notion of $(\bbp,\delta)$-dense sequences to $(\cP,\Delta)$-dense sequences to construct a position-estimation code for the $t$-localized-deletion model.

In this paper, we continue the study of $(\le t)$-burst-deletion correcting codes and $t$-localized-deletion correcting codes. Our contributions are:
\begin{itemize}
    \item a new construction of $q$-ary $(\le t)$-burst-deletion correcting codes with redundancy $\log n+(t-1)\log\log n+O(1)$, for any fixed $q\ge 2$ and $2\le t<2q$;
    \item a new construction of $q$-ary $t$-localized-deletion correcting codes with redundancy $\log n+(t-1)\log\log n+O(1)$, for any fixed $q\ge 2$ and $2\le t<2q$;
    \item an efficient encoding algorithm which encodes an arbitrary length-$(n-2)$ sequence into a length-$n$ sequence whose differential sequence is strong-$(\ell,\epsilon)$-locally-balanced.
\end{itemize}

\subsection{Comparison}
The first step of our error-correction algorithms is also to estimate the positions of deleted symbols. However, the position-estimation code is different from previous constructions. We select our codewords from the set of sequences whose differential sequences are strong-$(\ell,\epsilon)$-locally-balanced. Then we impose a VT-type constraint and an $L_1$-weight constraint on the differential sequences to give single position-estimation code. In the following, we compare our results with previous ones and explain how the improvement is gained.

\vspace{3pt}
\subsubsection{correcting burst-deletion} The position-estimation codes in \cite{Lenz2020ISIT} (also, \cite{Song2023IT,Song2025Entropy}, \cite[Section V]{Shuche2024IT},\cite[Section VII-C]{Yubo2025IT}) and this paper can approximately determine error positions within a window of length $O(\log n)$, where $n$ is the length of codewords. Our position-estimation code can also correct a single-deletion while the position-estimation codes in those works cannot. This is why the redundancy of our code is smaller than that of the code in \cite{Yubo2025IT} by $\log\log n$. In addition, our position-estimation code handles $q$-ary sequence directly while the one in the construction in \cite{Yubo2025IT} handles binary sequences associated with $q$-ary codewords. This is why our results holds for all $q$ but the result in \cite{Yubo2025IT} only holds for even $q$. It is worth noting that our construction requires $t<2q$ while constructions in previous works do not have this constraint. At last, the proof for the correctness of our position-estimation code is simpler. For details, see the proofs of \Cref{thm_generalburst}, \cite[Lemma 2]{Lenz2020ISIT} and \cite[Lemma 4]{Song2025Entropy}.

Our position-estimation code has some similarities with the one in \cite[Lemma 9]{Ryan2018IT}. The code there imposes a VT-type constraint and the strongly local balance property on binary codewords, which results in their construction being only applicable to the case where the burst-length is odd. Our code imposes both constraints on differential sequences of codewords. This allows for a wider range of $q$ and $t$. It is not difficult to verify that the construction in \cite[Lemma 9]{Ryan2018IT} can be generalized to larger alphabets to give a $q$-ary position-estimation code. However, this code cannot correct a single-deletion when $q>2$ and hence can only lead to $(\le t)$-burst-deletion codes with redundancy $\log n+t\log\log n+O(1)$.

\vspace{3pt}
\subsubsection{correcting localized deletions} The position-estimation code in \cite{RawadBitar2021isit} (and \cite{Yubo2025IT}) locates errors within a window of length $O\parenv{(\log n)^2}$. Our position-estimation code can locate errors within a window of length $O(\log n)$. Together with the fact that our position-estimation code corrects a single-deletion, the redundancy of our code improves that of the code in \cite{Yubo2025IT} by $(t+1)\log\log n$.

\vspace{3pt}
\subsubsection{encoding into strongly locally balanced differential sequences} In \cite[Section VI]{Yubosun2025IT}, Sun and Ge introduced an algorithm which efficiently encodes an arbitrary binary sequence into a strong-$(\ell,\epsilon)$-locally-balanced sequence with only one redundant bit. In this work, we first generalize this encoding idea to larger alphabets. Then, we show how to modify this algorithm to encode a sequence into another sequence whose differential sequence is strong-$(\ell,\epsilon)$-locally-balanced.

To the best of our knowledge, no such an algorithm exists in previous works. Although one can apply the generic framework in \cite{BarLev202304} to design an encoding algorithm with only one redundant symbol, the algorithm can only guarantee averagely efficient running time. In contrast, our algorithm can efficiently give the output for \emph{each} input.

\vspace{3pt}
The rest of this paper is organized as follows. In \Cref{sec_pre}, some necessary notations and definitions are given. In \Cref{sec_aux}, we introduce auxiliary results that are needed in our constructions. In \Cref{sec_burst,sec_localized}, we construct $(\le t)$-burst-deletion correcting codes and $t$-localized-deletion correcting codes, respectively. \Cref{sec_algorithms} is devoted to presenting the encoding algorithm mentioned above. Lastly, \Cref{sec_conclusion} concludes this paper.

\section{Preliminary}\label{sec_pre}
%%%%%%%%%%%%%%%%%%%%%%%%%%%%%%%%%%%%%%%%%%
For an integer $q\ge 2$, define $\Sigma_q\triangleq\mathset{0,1,\ldots,q-1}$ to be the $q$-ary alphabet. For $n\ge0$, let $\Sigma_q^n$ denote the set of all $q$-ary sequences of length $n$. The unique length-$0$ sequence is the empty sequence. For a sequence $\bfx\in\Sigma_q^n$, its $i$-th entry is referred to as $x_i$ and its length is denoted by $\abs{\bfx}$. The \emph{concatenation} of two sequences $\bfx$ and $\bby$ is denoted by $\bfx\bby$. For integers $m\le n$, let $\sparenv{m,n}$ denote the set $\mathset{m,m+1,\ldots,n}$. For simplicity, we abbreviate $[1,n]$ as $[n]$. For a sequence $\bfx\in\Sigma_q^n$ and a subset $I=\mathset{i_1,\ldots,i_k}\subseteq[n]$ where $i_1<\cdots<i_k$, denote $\bfx_{I}=x_{i_1}\cdots x_{i_k}$ and call $\bfx_{I}$ a \emph{subsequence} of $\bfx$. In particular, when $I=[a,b]\subseteq[n]$, $\bfx_{I}$ is called a \emph{substring} of $\bfx$. We also regard $\bfx_{[a,a-1]}$ as the empty sequence, for any $a\ge 1$.

We say that sequence $\bby$ results from $\bfx$ by a $(\le t)$-burst-deletion, if it is obtained from $\bfx$ by deleting a substring of length \emph{at most} $t$. Let $\cD_{\le t}(\bfx)$ be the set of all such sequences. The set $\cD_{\le t}(\bfx)$ is called the $(\le t)$-burst-deletion ball centered at $\bfx$.
\begin{definition}
A non-empty subset $\cC\subseteq\Sigma_q^n$ is called a $(\le t)$-burst-deletion correcting code, if $\cD_{\le t}(\bfx)\cap\cD_{\le t}(\bby)=\emptyset$ for any distinct $\bfx,\bby\in\cC$. 
\end{definition}

In addition to $(\le t)$-burst-deletion, we also study a more general error, called a $t$-localized-deletion. A sequence $\bfx$ is said to undergo a $t$-localized-deletion, if it experiences \emph{at most} $t$ deletions confined to a single window of length $t$ and the location of this window is not known a priori.\footnote{Some authors assume that the length of the window is at most $t$. There is no fundamental distinction between these two assumptions.} By definition, a $2$-localized-deletion is the same as a burst of at most $2$ deletions. Therefore, when referring to $t$-localized-deletion, we always assume $t\ge3$.
The precise definition is structured as follows.
\begin{definition}\label{dfn_localized}
 Let $t\ge3$ and $2\le t^\prime\le t$. Let $\bfx\in\Sigma_q^{n}$. We say that sequence $\bby\in\Sigma_q^{n-t^\prime}$ is obtained from $\bfx$ by a $t$-localized-deletion, if there are  positive integers $t_1,\ldots,t_{k}$ and indices $1\le i_1<\cdots<i_{k}\le n-t_k+1$ satisfying: 1) $i_k-i_1\le t-t_k$ (\emph{window constraint}); 2) $i_{j}-i_{j-1}>t_{j-1}$ for all $1<j\le k$ (\emph{non-adjacent condition}); 3) $t_1+\cdots+t_k=t^\prime$, such that $\bby$ is obtained from $\bfx$ by deleting the $k$ substrings: $\bfx_{\sparenv{i_j,i_j+t_j-1}}$ ( $j=1,\ldots,k$).
\end{definition}

\begin{remark}
    We briefly explain conditions 1), 2) and 3) in the above definition. Condition 1) ensures that all deleted symbols are located in a window of length $t$. Condition 2) ensures that any two of the $k$ deleted substrings are not adjacent, since otherwise they can be treated as single substring. Condition 3) says that exactly $t^\prime$ symbols are deleted.
\end{remark}

Let $\cD_{t}^{\rm{loc}}(\bfx)$ be the set of sequences which results from $\bfx$ by a $t$-localized-deletion.
\begin{definition}
A non-empty subset $\cC\subseteq\Sigma_q^n$ is called a $t$-localized-deletion correcting code, if $\cD_{t}^{\rm{loc}}(\bfx)\cap\cD_{t}^{\rm{loc}}(\bby)=\emptyset$ for any distinct $\bfx,\bby\in\cC$. 
\end{definition}

For any integer $m>1$ and real number $x\ge 1$, denote by $\log_m x$ the logarithm of $x$ to the base $m$. If $m=2$, we omit $m$ in this notation. The \emph{redundancy} of a code $\cC\subseteq\Sigma_q^n$, denoted by $\rho(\cC)$, is defined to be $\log\parenv{q^n/\abs{\cC}}$, where $\abs{\cC}$ is the cardinality of $\cC$.  Throughout this paper, it is assumed that $q$ and $t$ are fixed constants and the code-length $n$ is sufficiently large. For a set $\cS$ of parameters, let $O_{\cS}(1)$ denote a positive number only dependent on parameters in $\cS$.

As mentioned in the \Cref{sec_introduction}, the redundancy of a $(\le t)$-burst-deletion or a $t$-localized-deletion correcting code is lower bounded by $\log n+O_{q,t}(1)$. We will give constructions of $(\le t)$-burst-deletion correcting codes and $t$-localized-deletion correcting codes with redundancy $\log n+(t-1)\log\log n+O_{q,t}(1)$ in \Cref{sec_burst,sec_localized}, respectively.
We will prove the correctness of constructed codes by showing an error-correcting algorithm with $O\parenv{n\log n}$ time complexity. Briefly speaking, the algorithm operates in two steps: firstly roughly locating errors within a short substring of codewords and then correcting errors in the short substring.

In the next section, we introduce some results which will serve as critical components in our constructions. Before that, we define two functions on sequences. For any sequence $\bfx\in\Sigma_q^n$, define $\vt(\bfx)\triangleq\sum_{i=1}^{n}ix_i$ and $\Sum{\bfx}\triangleq\sum_{i=1}^nx_i$. We also call $\Sum{\bfx}$ the \emph{$L_1$-weight} of $\bfx$.  

\section{Auxiliary Results}\label{sec_aux}
%%%%%%%%%%%%%%%%%%%%%%%%%%%%%%%%%%%%%%%%%%%%%%%%%
In this section, we introduce some auxiliary results and explain along the way the role each of them plays.
\subsection{differential sequences}\label{sec_diff}
The \emph{differential} sequence of sequence $\bfx\in\Sigma_q^n$ is defined to be $\bby=\dvt(\bfx)$ where $y_i\triangleq (x_i-x_{i+1})\pmod{q}$ for all $1\le i< n$ and $y_{n}\triangleq x_n$. 

It was shown in \cite{Shuche2024IT} that for any $0\le a<2n$, the binary code $\mathset{\bfx\in\Sigma_2^n:\vt(\dvt(\bfx))\equiv a\pmod{2n}}$ is a $(\le 2)$-burst-deletion correcting code. One may ask if this conclusion can be generalized to larger alphabets. Putting it in another way, when $q\ge3$, if there exists some positive integer $N$, such that the following set
$$
\cC_{\dvt}\parenv{q,n;N}\triangleq\mathset{\bfx\in\Sigma_q^n:\vt(\dvt(\bfx))\equiv a\pmod{N}}
$$
is a $(\le 2)$-burst-deletion correcting code? Unfortunately, the answer is negative, as demonstrated in the next example.
\begin{example}
    Suppose that $q=3$ and $n=4$. Let $\bfx=0200,\bbz=0110\in\Sigma_q^n$. Then we have $\vt(\dvt(\bfx))=\vt(\dvt(\bbz))$. On the other hand, it is easy to see that $\bfx_{[4]\setminus\{2,3\}}=\bbz_{[4]\setminus\{2,3\}}$.
\end{example}

Therefore, it is impossible to guarantee the capability of correcting two adjacent deletions by imposing only the $\vt$ constraint on the differential sequences. Now it is interesting to ask the following question.
\begin{question}\label{question1}
    Can one introduce additional $o\parenv{\log n}$ redundant bits to the code $\cC_{\dvt}\parenv{q,n;N}$ to get a code which can correct a burst of at most two deletions?
\end{question}

Our construction in \Cref{thm_generalburst} shows that the answer to \Cref{question1} is positive. 

For any $\alpha,\beta\in\Sigma_q$, define $\alpha\oplus \beta\triangleq(\alpha+\beta)\pmod{q}$. Suppose that $\bfx\in\Sigma_q^n$ and $\bfx^\prime$ results from $\bfx$ by deleting the $i$-th entry $x_i$. Let $\bby\in\dvt(\bfx)$ and $\bby^\prime\in\dvt\parenv{\bfx^\prime}$.
It was proved in \cite{Tuan2024IT} that
\begin{itemize}
    \item If $2\le i\le n$, then $\bby^\prime$ is obtained from $\bby$ by replacing $y_{i-1}y_{i}$ with one symbol $y_{i-1}\oplus y_i$.
    \item If $i=1$, then $\bby^\prime$ is obtained from $\bby$ by deleting $y_1$.
\end{itemize}

Therefore, for uniform arguments, instead of $\dvt(\bfx)$, we will use a variant of the differential sequence: $\psi(\bfx)$, which is defined as $\psi(\bfx)\triangleq\dvt(0\bfx)$. By abuse of terminology, we also call it the differential sequence of $\bfx$. Clearly, if we define $x_0=x_{n+1}=0$, then $\psi(\bfx)_i=(x_{i-1}-x_{i})\pmod{q}$. In addition, it is easy to see that $\psi(\cdot)$ is invertible. The following observation is straightforward.
\begin{observation}\label{obs_sumpsi}
\begin{enumerate}[$(i)$]
    \item Suppose that $\bfx^\prime$ is obtained from $\bfx$ by a deleting $x_i$ for some $1\le i\le n$. Let $\bby=\psi(\bfx)$ and $\bby^\prime=\psi\parenv{\bfx^\prime}$. Then $\bby^\prime$ is obtained from $\bby$ by replacing $y_{i}y_{i+1}$ with one symbol $y_{i}\oplus y_{i+1}$.
    \item It holds that $\Sum{\psi(\bfx)}\equiv 0\pmod{q}$ for any $\bfx$. When acting on sequences of length $n$, the mapping $\psi(\cdot)$ is a bijection between $\Sigma_q^n$ and the set
\begin{equation}\label{eq_zerosumset}
\mathset{\bby\in\Sigma_q^{n+1}:\Sum{\bby}\equiv0\pmod{q}}.
\end{equation}
\end{enumerate}
\end{observation}
\begin{IEEEproof}
    Conclusion (i) is a straightforward corollary of \cite[Lemma 1]{Tuan2024IT}. We prove conclusion (ii) here. By definition, we have $\psi(\bfx)_i=(x_{i-1}-x_i)\pmod{q}$ for all $1\le i\le n+1$ where $x_0=x_{n+1}=0$. Then it follows that $\Sum{\psi(\bfx)}\equiv\sum_{i=1}^{n+1}(x_{i-1}-x_{i})\pmod{q}\equiv 0\pmod{q}$. On the other hand, it is easy to verify that $x_i=\sum_{k=i+1}^{n+1}\psi(\bfx)_k\pmod{q}$ for $1\le i\le n$. Therefore, $\psi$ is a bijection between $\Sigma_q^n$ and the set in \eqref{eq_zerosumset}.
\end{IEEEproof}

Let $N\ge (n+1)q$. For any $0\le a<N$, define
\begin{equation}\label{eq_codepsi}
  \cC_{\psi}(q,n;N)\triangleq\mathset{\bfx\in\Sigma_q^n:\vt\parenv{\psi(\bfx)}\equiv a\pmod{N}}.
\end{equation}
By presenting a linear-time error-correcting algorithm, the proof of \cite[Theorem 3]{Tuan2024IT} established that $\cC_{\psi}$ is a single-deletion correcting code if $N=mq$ whenever $m\ge n+1$. However, the proof of \cite[Theorem 3]{Tuan2024IT} works only when $q\mid N$. This is because the error-correcting algorithm presented there first determines the deleted symbol by applying \cite[Lemma 2]{Tuan2024IT}. However, this step is unnecessary. Here, we provide a modified proof which is valid for any $N\ge(n+1)q$. Beyond its independent interest, this proof inspires our constructions in \Cref{thm_generalburst,thm_localized}.
\begin{lemma}\label{lem_singledeletionpsi}
    Let $q,n\ge 2$. For any $N\ge(n+1)q$ and $0\le a<N$, the code $\cC_{\psi}(q,n;N)$ can correct deletion of one symbol.
\end{lemma}
\begin{IEEEproof}
Suppose that $\bfx\in\cC_{\psi}(q,n;N)$ and $\bfx^\prime$ results from $\bfx$ by deleting $x_i$. Let $\bby=\psi(\bfx)$ and $\bby^\prime=\psi\parenv{\bfx^\prime}$. By \Cref{obs_sumpsi} (ii), it suffices to recover $\bby$ from $\bby^\prime$.

According to \Cref{obs_sumpsi} (i), $\bby^\prime$ is obtained from $\bby$ by replacing $y_{i}y_{i+1}$ with $y_{i}\oplus y_{i+1}$. Let $\Delta=\vt(\bby)-\vt\parenv{\bby^\prime}$. Then we have
\begin{equation}\label{eq_singledeletiondelta}
    \Delta=i\Delta_{sum}+y_{i+1}+\Sum{\bby^\prime_{[i+1,n]}},
\end{equation}
where $\Delta_{sum}=y_{i}+y_{i+1}-y_i\oplus y_{i+1}$. By definition, it holds that $\Delta_{sum}=\mathset{0,q}$. If $\Delta_{sum}=0$, then $y_{i+1}\le y_i\oplus y_{i+1}=y^\prime_{i}$ and it follows from \Cref{eq_singledeletiondelta} that $0\le\Delta\le\Sum{\bby^\prime}\le n(q-1)$. If $\Delta_{sum}=q$, then $y^\prime_{i}<y_{i+1}$ and it follows from \Cref{eq_singledeletiondelta} that $\Delta>\Sum{\bby^\prime}+q$. In addition, when $\Delta_{sum}=q$, $\Delta$ increases with $i$. This implies that $\Delta\le nq+q-1=(n+1)q-1$. Therefore, we can obtain the value of $\Delta$ by computing $\parenv{a-\vt\parenv{\bby^\prime}}\pmod{N}$. Then comparing $\Delta$ and $\Sum{\bby^\prime}$, we can know whether $\Delta_{sum}=0$ or $q$.

Now we show how to recover $\bby$ from $\bby^\prime$ when given $\Delta$ and $\Delta_{sum}$. Scan $\bby^\prime$ right-to-left and find the largest $j$ such that there are $\alpha,\beta\in\Sigma_q$ such that 
$y^\prime_j=\alpha+\beta-\Delta_{sum}$ and
\begin{equation}\label{eq_singledelta2}
\Delta=j\Delta_{sum}+\beta+\Sum{\bby^\prime_{[j+1,n]}}.
\end{equation}
By \Cref{eq_singledeletiondelta}, such index $j$ exists and $i\le j$. In addition, such $j$ can be found in $O(n)$ time. Let $\bbz$ be the sequence obtained by replacing $y^\prime_j$ with $\alpha\beta$. The following claim completes the proof of this lemma.
\begin{clm}
    It holds that $\bbz=\bby$. 
\end{clm}
\textit{Proof of Claim:}  Combining \Cref{eq_singledeletiondelta,eq_singledelta2}, we obtain
\begin{equation}\label{eq_singleburstdifference}
    (j-i)\Delta_{sum}=\Sum{\bby^\prime_{[i+1,j]}}+y_{i+1}-\beta.
\end{equation}
If $i=j$, \Cref{eq_singleburstdifference} implies that $\beta=y_{i+1}$ and $\alpha=y_i$. Then the conclusion follows. In the rest, assume that $i<j$.

We first consider the case when $\Delta_{sum}=0$. In this case, we have $\beta\le y^\prime_j$. Then it follows from \Cref{eq_singleburstdifference} that $0=\Sum{\bby^\prime_{[i+1,j-1]}}+y_{i+1}+y^\prime_j-\beta\ge0$. Therefore, it must be that $y_{i+1}=y_{i+1}^\prime=\cdots=y_{j-1}^\prime=\alpha=0$ and $y_j^\prime=\beta$. Since $\bby$ is obtained from $\bby^\prime$ by replacing $y_i^\prime$ with $y_iy_{i+1}=y_i^\prime0$, we have $\bbz=\bby$.

Next we consider the case when $\Delta_{sum}=q$. In this case, we have $y_j^\prime<\beta$. Then \Cref{eq_singleburstdifference} implies that $(j-i)q=\Sum{\bby^\prime_{[i+1,j-1]}}+y_{i+1}+y^\prime_j-\beta \le (j-i)(q-1)+y^\prime_j-\beta<(j-i)q$, which is a contradiction.
\end{IEEEproof}

\vspace{5pt}
One might ask if the above proof works when dealing with a burst of $t$ deletions, where $t\ge2$. Suppose that $\bfx^\prime$ results from $\bfx\in\Sigma_q^n$ by deleting the substring $\bfx_{[i,i+t-1]}$. As in the above proof, let $\bby=\psi(\bfx)$ and $\bby^\prime=\psi\parenv{\bfx^\prime}$. Then $\bby^\prime$ is obtained from $\bby$ by replacing $\bby_{[i,i+t]}$ with $\oplus_{k=i}^{i+t}y_k$. Let $\Delta=\vt(\bby)-\vt\parenv{\bby^\prime}$ and $\Delta_{sum}=\Sum{\bby_{[i,i+t]}}-\oplus_{k=i}^{i+t}y_k$. Then we have $\Delta=i\Delta_{sum}+\sigma^{(i)}+t\sum_{k=i+1}^{n+1-t}y_k^\prime$ where $0\le\sigma^{(i)}=\sum_{k=i+1}^{i+t}(k-i)y_k\le\binom{t+1}{2}(q-1)$ (see \Cref{eq_qburstdeltai}). Starting from the end of $\bby^\prime$, find the largest $j$ such that $\Delta=j\Delta_{sum}+\sigma^{(j)}+t\sum_{k=j+1}^{n+1-t}y_k^\prime$ for some $0\le\sigma^{(j)}\le\binom{t+1}{2}(q-1)$.
    
In general, we can not recover $\bby$ by replacing $y^\prime_j$ with arbitrary $\alpha_1\cdots\alpha_{t+1}\in\Sigma_q^t$ where $y^\prime_j=\alpha_1+\cdots+\alpha_{t+1}-\Delta_{sum}$ and $\sum_{k=2}^{t+1}(k-1)\alpha_k=\sigma^{(j)}$. To see this, let $\bfx=0200\in\Sigma_3^4$ and $\bfx^\prime=00$. Then $i=t=2$, $\bby=01200$, $\bby^\prime=000$, $\Delta=8$, $\Delta_{sum}=3$ and $\sigma^{(i)}=2$. By previous discussion, we have $j=2$ and $\sigma^{(j)}=2$. Let $\alpha_1\alpha_2\alpha_3=201$. Then $\alpha_1+\alpha_2+\alpha_3-\Delta_{sum}=0=y_j^\prime$ and $\alpha_2+2\alpha_3=2=\sigma^{(j)}$. However, if replacing $y^\prime_2$ with $201$, we will get $02010$ which is not $\bby$.

On the other hand, by the choice of $j$, it holds that
\begin{equation}\label{eq_analysis1}
  (j-i)\Delta_{sum}=t\sum_{k=i+1}^{j}y_{k}^\prime+\sigma^{(i)}-\sigma^{(j)}=t\sum_{k=i+1+t}^{j+t}y_{k}+\sigma^{(i)}-\sigma^{(j)}.
\end{equation}
If $\sum_{k=i+1+t}^{j+t}y_{k}$ is not too small or not too large, this equation never holds (see \Cref{clm_generalburst}). It turns out that $\bby$ should satisfy certain locally-balanced property as described below.

\subsection{strong-locally-balanced sequences}
Given $\ell\le n$ and $0<\epsilon<1/2$, a sequence $\bfx\in\Sigma_2^n$ is called a \emph{strong-$(\ell,\epsilon)$-locally-balanced} sequence, if any substring of $\bfx$ of length $\ell^\prime\ge\ell$ has Hamming weight in the interval $\sparenv{\parenv{\frac{1}{2}-\epsilon}\ell^\prime,\parenv{\frac{1}{2}+\epsilon}\ell^\prime}$.

In 2018, Gabrys \textit{et al} \cite{Ryan2018IT} introduced the concept of strong-$(\ell,\epsilon)$-locally-balanced sequences (over the binary alphabet) to construct burst-deletion-correcting codes. For a fixed $t\ge 5$, by imposing a VT-type constraint and other constraints on strong-$(\ell,\epsilon)$-locally-balanced sequences, they constructed a binary code which can correct single burst-deletion, as long as the length of the burst is \emph{odd} and at most $t$ \cite[Theorem 2]{Ryan2018IT}. The redundancy of their code is $\log n+O\parenv{t^2\log\log n}$. Binary strong-$(\ell,\epsilon)$-locally-balanced sequences also play critical roles in constructions of other codes \cite{Yubo2024IT,Yubosun2025IT}.

Notice that when $q=2$, the $L_1$-weight is nothing but the Hamming weight. Therefore, the notion of strong-$(\ell,\epsilon)$-locally-balanced sequences can be naturally generalized to larger alphabets.
\begin{definition}
    Let $q,n>1$ be integers. Let $\ell\le n$ and $0<\epsilon<(q-1)/2$. A sequence $\bfx\in\Sigma_q^n$ is called a \emph{strong-$(\ell,\epsilon)$-locally-balanced} sequence, if any substring of $\bfx$ of length $\ell^\prime\ge\ell$ has $L_1$-weight in the interval $\sparenv{\parenv{\frac{q-1}{2}-\epsilon}\ell^\prime,\parenv{\frac{q+1}{2}+\epsilon}\ell^\prime}$.
\end{definition}

 The first conclusion in the following lemma generalizes \cite[Claim 4]{Ryan2018IT}.
\begin{lemma}\label{lem_slocalbalanced}
    Suppose that $q,n\ge 2$ and $s\ge 1$. If $\ell\ge\frac{(q-1)^2}{\epsilon^2}\log_{\mathsf{e}}\parenv{2n\sqrt{s}}$, then there are at least $q^n\parenv{1-\frac{1}{2s}}$ $q$-ary strong-$(\ell,\epsilon)$-locally-balanced sequences of length $n$. In particular, when $\ell\ge\frac{(q-1)^2}{\epsilon^2}\log_{\mathsf{e}}\parenv{2(n+1)\sqrt{q}}$, there are at least $q^n/2$ sequences $\bfx\in\Sigma_q^n$ with $\psi(\bfx)$ being strong-$(\ell,\epsilon)$-locally-balanced.
\end{lemma}
\begin{IEEEproof}
    Select uniformly and randomly a sequence $\bfx\in\Sigma_q^n$. Then for any $1\le m\le n$, we have $\mathsf{E}(\Sum{\bfx_{[1,m]}})=\frac{q-1}{2}m$. By the Hoeffding’s inequality, we get
    \begin{equation*}
        \mathsf{P}\parenv{\abs{\Sum{\bfx_{\sparenv{1,m}}}-\frac{q-1}{2}m}>\epsilon m}\le 2\mathsf{e}^{-\frac{2\epsilon^2 m}{(q-1)^2}}.
    \end{equation*}
Therefore, it holds that
\begin{equation*}
    \mathsf{P}\parenv{\bfx\text{ is not strong-}(\ell,\epsilon)\text{-locally-balanced}}\le 2n^2\mathsf{e}^{-\frac{2\epsilon^2\ell}{(q-1)^2}}.
\end{equation*}
This implies that the number of strong-$(\ell,\epsilon)$-locally-balanced sequences of length $n$ is at least $q^n\parenv{1-2n^2\mathsf{e}^{-\frac{2\epsilon^2\ell}{(q-1)^2}}}$.
If $\ell\ge\frac{(q-1)^2}{\epsilon^2}\log_{\mathsf{e}}\parenv{2n\sqrt{s}}$, we have $2n^2\mathsf{e}^{-\frac{2\epsilon^2\ell}{(q-1)^2}}\le \frac{1}{2s}$. Therefore, the number of strong-$(\ell,\epsilon)$-locally-balanced sequences of length $n$ is at least $q^n\parenv{1-\frac{1}{2s}}$.

In particular, when $\ell\ge\frac{(q-1)^2}{\epsilon^2}\log_{\mathsf{e}}\parenv{2(n+1)\sqrt{q}}$, there are at least $q^{n+1}\parenv{1-\frac{1}{2q}}$ strong-$(\ell,\epsilon)$-locally-balanced sequences of length $n+1$. On the other hand, there are exactly $q^n$ sequences $\bby\in\Sigma_q^{n+1}$ with $\Sum{\bby}\equiv 0\pmod{q}$. Therefore, by \Cref{obs_sumpsi} (ii) and the inclusion-exclusion principle, there are at least $q^n/2$ sequences $\bfx\in\Sigma_q^n$ such that $\psi(\bfx)$ is strong-$(\ell,\epsilon)$-locally-balanced.
\end{IEEEproof}

As previously analyzed, if $\bby=\psi(\bfx)$ is strongly locally-balanced and \Cref{eq_analysis1} holds,  $j-i$ could not be too large. Then all positions of errors are determined within a short substring of $\bby$. Now it suffices to correct errors in this short substring. This necessitates the use of $(t_1,t_2)$-burst-error-correcting codes introduced next.

\subsection{$(t_1,t_2)$-burst-error correcting codes}
Let $1\le t_2\le t_1\le n$ be three integers. We say that sequence $\bby\in\Sigma_q^{n-t_1+t_2}$ results from $\bfx\in\Sigma_q^n$ by a $(t_1,t_2)$-burst-error, if there is some $i$ such that $\bby$ is obtained from $\bfx$ by replacing the substring $\bfx_{\sparenv{i,i+t_1-1}}$ with a $q$-ary sequence of length $t_2$.

The study of this kind of error was initiated in \cite{Schoeny2017it} and was investigated in-depth in subsequent two works \cite{Ziyang2023IT,Yubo2025IT}. When $q$, $t_1$ and $t_2$ are fixed, Sun \textit{et al} constructed a $(t_1,t_2)$-burst-error correcting code with redundancy $\log n+O_{q,t_1,t_2}(1)$, which is optimal up to a constant.

For each $t^\prime\in [1,t]$, let $\cC_{t^\prime}$ be a $(t,t-t^\prime)$-burst-error correcting code.
By definition, the intersection $\cap_{t^\prime=1}^{t} \cC_{t^\prime}$ can correct a $t$-localized-deletion. Since we just have to correct errors in short a substring, a $P$-bounded $(t_1,t_2)$-burst-error correcting code is needed.
\begin{definition}[$P$-bounded code]
    A code $\cC\subseteq\Sigma_q^n$ is said to be a $P$-bounded $(t_1,t_2)$-burst-error correcting code, where $P\le n$, if it can correct a $(t_1,t_2)$-burst-error when it is known that all errors occur in a known interval of length $P$.
\end{definition}

\begin{lemma}\cite[Corollary 3]{Yubo2025IT}\label{lem_Pboundedsingleburst}
    Let $q\ge 2$ and $t_1\ge t_2\ge 2$ be three fixed integers. For any $P\le n$, there is a function $f_{P,t_1,t_2}:\Sigma_q^n:\rightarrow\{0,1\}^{\log P+O_{q,t_1,t_2}(1)}$, computable in $O(n)$ time, such that for any $a\in\{0,1\}^{\log P+O_{q,t_1,t_2}(1)}$, the code
    \begin{equation*}
        \mathset{\bfx\in\Sigma_q^n:f_{P,t_1,t_2}(\bfx)=a}
    \end{equation*}
    is a $P$-bounded $(t_1,t_2)$-burst-error correcting code and it can correct a $(t_1,t_2)$-burst-error in $O(nP)$ time by brute-force.
\end{lemma}

\subsection{good triples}
Our constructions in \Cref{sec_burst,sec_localized} are valid if and only if $t<2q$. In this case, there always exists some $0<\epsilon<\frac{q-1}{2}$ such that the triple $(q,t,\epsilon)$ satisfies a good property, which is critical in proofs of \Cref{thm_generalburst,thm_localized}. We call such triples good triples. The formal definition is given below.
\begin{definition}
    Let $q,t\ge 2$ be integers and $0<\epsilon<\min\mathset{\frac{q}{2t},\frac{1}{2}}$. If for any $2\le t^\prime\le t$, there is an integer $s_{t^\prime}\in\sparenv{1,t^\prime}$ such that
\begin{equation}\label{eq_goodtriple1}
\frac{t^\prime}{2}-\frac{(1-2\epsilon)t^\prime}{2q}<s_{t^\prime}<\frac{t^\prime}{2}+1-\frac{(1+2\epsilon)t^\prime}{2q},
\end{equation}
we call $\parenv{q,t,\epsilon}$ a \emph{good triple}. 
\end{definition}
\begin{remark}\label{rmk_goodtriple}
\begin{itemize}
    \item Suppse that $\parenv{q,t,\epsilon}$ is a good triple. It is necessary that $\frac{t^\prime}{2}-\frac{(1-2\epsilon)t^\prime}{2q}<\frac{t^\prime}{2}+1-\frac{(1+2\epsilon)t^\prime}{2q}$, or equivalently, $\epsilon<\frac{q}{2t^\prime}$, for all $2\le t^\prime\le t$. In particular, we have $\epsilon<\frac{q}{2t}$. If $\epsilon\ge\frac{1}{2}$, it follows from (\ref{eq_goodtriple1}) that $\frac{t^\prime}{2}<s_{t^\prime}<\frac{t^\prime}{2}+1$ for all $2\le t^\prime\le t$. However, this is impossible when $t^\prime=2$. This discussion explains why it is required that $\epsilon<\min\mathset{\frac{q}{2t},\frac{1}{2}}$ in the above definition.
    \item Since $\frac{t^\prime}{2}+1-\frac{(1+2\epsilon)t^\prime}{2q}-\parenv{\frac{t^\prime}{2}-\frac{(1-2\epsilon)t^\prime}{2q}}<1-\frac{2\epsilon t^\prime}{q}<1$, if $(q,t,\epsilon)$ is a good triple, the integer $s_{t^\prime}$ is unique for each $t^\prime$.
\end{itemize}
\end{remark}

In \Cref{lem_goodtriple}, we determine the set of all good triples. The following observation will be helpful in deriving results in \Cref{lem_goodtriple}.
\begin{observation}\label{obs_goortriple1}
    Let $q, t^\prime\ge2$ be integers and $0<\epsilon<\min\mathset{\frac{q}{2t^\prime},\frac{1}{2}}$. Let
    \begin{equation}\label{eq_openinterval}
        I_{t^\prime}=\parenv{\frac{t^\prime}{2}-\frac{(1-2\epsilon)t^\prime}{2q},\frac{t^\prime}{2}+1-\frac{(1+2\epsilon)t^\prime}{2q}}
    \end{equation}
    be an open interval on the line of real numbers.
    \begin{itemize}
        \item If $t^\prime$ is even, the interval $I_{t^\prime}$ contains an integer if and only if there is an integer $s$ with $\frac{(1+2\epsilon)t^\prime}{2q}-1<s<\frac{(1-2\epsilon)t^\prime}{2q}$.
        \item If $t^\prime$ is odd, the interval $I_{t^\prime}$ contains an integer if and only if there is an integer $s$ with $\frac{(1+2\epsilon)t^\prime}{2q}-\frac{1}{2}<s<\frac{(1-2\epsilon)t^\prime}{2q}+\frac{1}{2}$.
    \end{itemize}
\end{observation}
\begin{IEEEproof}
    Let $s_{t^\prime}\in I_{t^\prime}$ be an integer. It must be that $s_{t^\prime}\le\frac{t^\prime+1}{2}$. When $t^\prime$ is even, suppose that $s_{t^\prime}=\frac{t^\prime}{2}-s$ where $s\ge0$ is an integer. It is easy to verify that $\frac{t^\prime}{2}-s\in I_{t^\prime}$ if and only if $\frac{(1+2\epsilon)t^\prime}{2q}-1<s<\frac{(1-2\epsilon)t^\prime}{2q}$. When $t^\prime$ is odd, suppose that $s_{t^\prime}=\frac{t^\prime+1}{2}-s$ where $s\ge0$ is an integer. It is easy to verify that $\frac{t^\prime+1}{2}-s\in I_{t^\prime}$ if and only if $\frac{(1+2\epsilon)t^\prime}{2q}-\frac{1}{2}<s<\frac{(1-2\epsilon)t^\prime}{2q}+\frac{1}{2}$.
\end{IEEEproof}
\begin{lemma}\label{lem_goodtriple}
    Let $q,t\ge2$ be integers and $0<\epsilon<\min\mathset{\frac{q}{2t},\frac{1}{2}}$. Let $\delta\in\{0,1\}$ be such that $\delta\equiv t\pmod{2}$. Define $t_1=t-1+\delta$ and $t_2=t-\delta$.
    Then $(q,t,\epsilon)$ is a good triple if and only if one of the following holds:
    \begin{enumerate}[$(i)$]
        \item $q>t\ge 2$ and $0<\epsilon<\min\mathset{\frac{q}{2t_1}-\frac{1}{2},\frac{q}{2t},\frac{1}{2}}$;
        \item $q$ is even, $t=q$ and $0<\epsilon<\frac{1}{2(q-1)}$; or $q$ is even, $q<t<2q$ and $0<\epsilon<\min\mathset{\frac{q}{2t},\frac{q}{t_2}-\frac{1}{2},\frac{1}{2(q+1)}}$.
    \end{enumerate}
    Furthermore, if $(q,t,\epsilon)$ is a good triple, let $s_{t^\prime}$ be the unique integer in $I_{t^\prime}$. Then $s_{t^\prime}=\ceilenv{t^\prime/2}$ when $t^\prime\le q$ and $s_{t^\prime}=\floorenv{t^\prime/2}$ when $t^\prime>q$.
\end{lemma}
\begin{IEEEproof}
For any $2\le t^\prime\le t$, let $I_{t^\prime}$ be the interval defined in (\ref{eq_openinterval}). To prove this lemma, it suffices to show that $I_{t^\prime}$ contains an integer for any $2\le t^\prime\le t$ if and only if (i) or (ii) is satisfied. In other words, it suffices to show that one of the two items in \Cref{obs_goortriple1} is satisfied if and only if (i) or (ii) is satisfied.

(i) Suppose that $q>t$. Then it is easy to see that $\frac{(1+2\epsilon)t^\prime}{2q}-1<0$, $\frac{(1-2\epsilon)t^\prime}{2q}>0$ for any $2\le t^\prime\le t$. Therefore, when $t^\prime$ is even, the interval $\parenv{\frac{(1+2\epsilon)t^\prime}{2q}-1,\frac{(1-2\epsilon)t^\prime}{2q}}$ always contains integer $0$. That is to say, the first item in \Cref{obs_goortriple1} is always satisfied and $s_{t^\prime}=t^\prime/2$. When $t^\prime$ is odd, since $0<\frac{(1-2\epsilon)t^\prime}{2q}+\frac{1}{2}<1-\epsilon$, the open interval $\parenv{\frac{(1+2\epsilon)t^\prime}{2q}-\frac{1}{2},\frac{(1-2\epsilon)t^\prime}{2q}+\frac{1}{2}}$ contains an integer if and only if $\frac{(1+2\epsilon)t^\prime}{2q}-\frac{1}{2}<0$, i.e., $\epsilon<\frac{q}{2t^\prime}-\frac{1}{2}$. Therefore, the second item in \Cref{obs_goortriple1} is satisfied if and only if $\epsilon<\frac{q}{2t_1}-\frac{1}{2}$. In this case, we have $s_{t^\prime}=(t^\prime+1)/2$.

(ii) Suppose that $t\ge q$. In this case, it is necessary that $t<2q$. Indeed, if $t\ge2q$, set $t^\prime=2q$. Since $(q,t,\epsilon)$ is a good triple, there is an integer in the interval $I_{t^\prime}$. Then it follows from \Cref{obs_goortriple1} that there is an integer $s$ satisfying $2\epsilon=\frac{(1+2\epsilon)t^\prime}{2q}-1<s<\frac{(1-2\epsilon)t^\prime}{2q}=1-2\epsilon$, which is impossible.

Next, we show that $q$ must be even. Since $t\ge q$, we can set $t^\prime=q$. Then $\frac{(1+2\epsilon)t^\prime}{2q}-\frac{1}{2}=\epsilon$ and $\frac{(1-2\epsilon)t^\prime}{2q}+\frac{1}{2}=1-\epsilon$. The interval $(\epsilon,1-\epsilon)$ does not contain integers. By \Cref{obs_goortriple1}, $q$ can not be odd.

It remains to show that one of the two items in \Cref{obs_goortriple1} is satisfied if and only if $\epsilon$ satisfies the bounds in (ii). If $2\le t^\prime<q$, it follows from (i) that $I_{t^\prime}$ contains an integer if and only if $\epsilon<\min\mathset{\frac{q}{2(q-1)}-\frac{1}{2},\frac{1}{2}}=\frac{1}{2(q-1)}$. Now suppose that $t^\prime=q+r$, where $0\le r\le t-q<q$. Since $q$ is even, it follows that $t^\prime\equiv r\pmod{2}$. When $r$ is even, since $0<\frac{(1-2\epsilon)t^\prime}{2q}<1-2\epsilon$, the first item in \Cref{obs_goortriple1} is satisfied if and only if $\frac{(1+2\epsilon)t^\prime}{2q}-1<0$, which is equivalent to $\epsilon<\frac{q}{q+r}-\frac{1}{2}$. It is easy to see that $s_{t^\prime}=t^\prime/2$ in this case. Let $r$ run through all even integers in $[0,t-q]$ and we obtain that $\epsilon<\frac{q}{t_2}-\frac{1}{2}$. This proves (ii) when $t=q$.

If $t>q$, $r$ can take odd values. When $r$ is odd, since $\frac{(1+2\epsilon)t^\prime}{2q}-\frac{1}{2}=\epsilon+\frac{(1+2\epsilon)r}{2q}>0$ and $\frac{(1-2\epsilon)t^\prime}{2q}+\frac{1}{2}<2$, the second item in \Cref{obs_goortriple1} is satisfied if and only if $\epsilon+\frac{(1+2\epsilon)r}{2q}<1$ and $\frac{(1-2\epsilon)t^\prime}{2q}+\frac{1}{2}>1$, which is equivalent to $\epsilon<\frac{r}{2(q+r)}$. It is easy to see that $s_{t^\prime}=(t^\prime-1)/2$ in this case. Let $r$ run through all odd integers in $[0,t-q]$ and we obtain that $\epsilon<\frac{1}{2(q+1)}$. Put everything together, we conclude that when $q$ is even and $q<t<2q$, one of the two items in \Cref{obs_goortriple1} is satisfied if and only if $0<\epsilon<\min\mathset{\frac{q}{2t},\frac{1}{2},\frac{q}{t_2}-\frac{1}{2},\frac{1}{2(q-1)},\frac{1}{2(q+1)}}=\min\mathset{\frac{q}{2t},\frac{q}{t_2}-\frac{1}{2},\frac{1}{2(q+1)}}$.
\end{IEEEproof}

\section{Burst-Deletion Correcting Codes}\label{sec_burst}
%%%%%%%%%%%%%%%%%%%%%%%%%%%%%%%%%%%%%%%%%%%%%%%
This section is devoted to the construction of $(\le t)$-burst-deletion correcting codes. Let $\parenv{q,t,\epsilon}$ be a good triple. By \Cref{rmk_goodtriple}, there is a unique integer in the interval $I_{t^\prime}$, for each $2\le t^\prime\le t$. Denote this integer by $s_{t^\prime}$. Define
\begin{equation}\label{eq_goodtriple2}
  M_{q,t,\epsilon}=\max\mathset{\frac{t^\prime(t^\prime+1)(q-1)}{2s_{t^\prime}q-t^\prime(q-1+2\epsilon)},\frac{t^\prime(t^\prime+1)(q-1)}{t^\prime(q-1-2\epsilon)-2(s_{t^\prime}-1)q}:2\le t^\prime\le t}.
\end{equation}
It follows from (\ref{eq_goodtriple1}) that $2s_{t^\prime}q-t^\prime(q-1+2\epsilon)>0$ and $t^\prime(q-1-2\epsilon)-2(s_{t^\prime}-1)q>0$ and thus, $M_{q,t,\epsilon}>0$. The next lemma completely determines $M_{q,t,\epsilon}$ for all good triples.

\begin{lemma}\label{lem_valueofM}
    Suppose that $(q,t,\epsilon)$ is a good triple. Let $t_1$ and $t_2$ be defined in \Cref{lem_goodtriple}.
    \begin{enumerate}[$(i)$]
        \item If $q>t$, then
        $$
        M_{q,t,\epsilon}=\max\mathset{\frac{(t_2+1)(q-1)}{1-2\epsilon},\frac{t_1(t_1+1)(q-1)}{q-(1+2\epsilon)t_1}}.
        $$
        \item If $t=q>2$,\footnote{The case $t=q=2$ has been studied in the literature.} then $M_{q,t,\epsilon}=\frac{q^3-2q^2+q}{1-2\epsilon(q-1)}$; if $2\le q<t<2q$, then $M_{q,t,\epsilon}=\frac{(q^2-1)(q+2)}{1-2\epsilon (q+1)}$.
    \end{enumerate}
\end{lemma}
\begin{IEEEproof}
    For $2\le t^\prime\le t$, denote
    $$
    f\parenv{t^\prime}=\frac{t^\prime(t^\prime+1)(q-1)}{2s_{t^\prime}q-t^\prime(q-1+2\epsilon)},\quad g\parenv{t^\prime}=\frac{t^\prime(t^\prime+1)(q-1)}{t^\prime(q-1-2\epsilon)-2(s_{t^\prime}-1)q}.
    $$

    (i) When $q>t\ge 2$, it follows from \Cref{lem_goodtriple} that $s_{t^\prime}=\ceilenv{t^\prime/2}$ for any $2\le t^\prime\le t$.
    If $t^\prime$ is even, then $f\parenv{t^\prime}=\frac{(q-1)(t^\prime+1)}{1-2\epsilon}$ and $g\parenv{t^\prime}=\frac{(q-1)t^\prime(t^\prime+1)}{2q-(1+2\epsilon)t^\prime}$. It is easy to verify that $f\parenv{t^\prime}>g\parenv{t^\prime}$ for all $t^\prime$ and $f\parenv{t^\prime}$ increases with $t^\prime$. Therefore, we have $\max\mathset{f\parenv{t^\prime},g\parenv{t^\prime}:t^\prime\text{ is even}}=f(t_2)$.
    If $t^\prime$ is odd, then $f\parenv{t^\prime}=\frac{(q-1)t^\prime(t^\prime+1)}{q+(1-2\epsilon)t^\prime}$ and $g\parenv{t^\prime}=\frac{(q-1)t^\prime(t^\prime+1)}{q-(1+2\epsilon)t^\prime}$. It is easy to see that $f\parenv{t^\prime}<g\parenv{t^\prime}$ for each $t^\prime$ and $g\parenv{t^\prime}$ increases with $t^\prime$. Therefore, we have $\max\mathset{f\parenv{t^\prime},g\parenv{t^\prime}:t^\prime\text{ is odd}}=g(t_1)$.
    The proof of (i) is completed by setting $M_{q,t,\epsilon}=\max\mathset{f(t_2),g(t_1)}$.

    (ii) When $q$ is even and $q\le t<2q$, it follows from \Cref{lem_goodtriple} that $s_{t^\prime}=\ceilenv{t^\prime/2}$ when $t^\prime\le q$ and $s_{t^\prime}=\floorenv{t^\prime/2}$ when $t^\prime>q$. By the argument in the proof of (i), we conclude that
    \begin{align*}
        &\max\mathset{f(t^\prime),g(t^\prime):2\le t^\prime\le q}=\max\mathset{f(q),g(q-1)}\\
        =&\max\mathset{\frac{q^2-1}{1-2\epsilon},\frac{q^3-2q^2+q}{1-2\epsilon(q-1)}}\\
        =&\begin{cases}
            \frac{3}{1-2\epsilon},\mbox{ if }q=2,\\
            \frac{q^3-2q^2+q}{1-2\epsilon(q-1)},\mbox{ if }q>2,
        \end{cases}
    \end{align*}
    and $\max\mathset{f\parenv{t^\prime},g\parenv{t^\prime}:q<t^\prime\le t\text{ is even}}=f(t_2)=\frac{(q-1)(t_2+1)}{1-2\epsilon}$. When $t^\prime>q$ is odd, we have $f\parenv{t^\prime}=\frac{(q-1)t^\prime(t^\prime+1)}{(1-2\epsilon)t^\prime-q}$ and $g\parenv{t^\prime}=\frac{(q-1)t^\prime(t^\prime+1)}{3q-(1+2\epsilon)t^\prime}$. It is clear that $f(t^\prime)>g(t^\prime)$ for all $q<t^\prime\le t<2q$. Since $f(t^\prime)$ decreases with $t^\prime$, we conclude that $\max\mathset{f(t^\prime),g(t^\prime):q<t^\prime\le t\text{ is odd}}=f(q+1)=\frac{(q^2-1)(q+2)}{1-2\epsilon (q+1)}$.

    Therefore, when $t=q>2$, we have $M_{q,t,\epsilon}=\frac{q^3-2q^2+q}{1-2\epsilon(q-1)}$; when $q=2$ and $t=3$, we have $M_{q,t,\epsilon}=\max\mathset{\frac{3}{1-2\epsilon},\frac{12}{1-6\epsilon}}=\frac{12}{1-6\epsilon}$; when $2<q<t<2q$, we have $M_{q,t,\epsilon}=\max\mathset{\frac{q^3-2q^2+q}{1-2\epsilon(q-1)},\frac{(q-1)(t_2+1)}{1-2\epsilon},\frac{(q^2-1)(q+2)}{1-2\epsilon (q+1)}}=\frac{(q^2-1)(q+2)}{1-2\epsilon (q+1)}$.
\end{IEEEproof}

Note that if $(q,t,\epsilon)$ is a good triple, it must be that $0<\epsilon<\frac{q-1}{2}$. Now we are ready to present the main result in this section.
\begin{theorem}\label{thm_generalburst}
    Let $\parenv{q,t,\epsilon}$ be a good triple, which is specified in \Cref{lem_goodtriple}. Let $M_{q,t,\epsilon}$ be given in \Cref{lem_valueofM}. Suppose that integer $n$ is sufficiently large such that $\ell=\ceilenv{\frac{(q-1)^2}{\epsilon^2}\log_{\mathsf{e}}\parenv{2(n+1)\sqrt{q}}}>M_{q,t,\epsilon}$. Let $P=\ell+t-1$ and $N\ge(nq+q-1)t$. For any $2\le t^\prime\le t$, let $f_{P,t^\prime,0}$ be the function in \Cref{lem_Pboundedsingleburst}. For any $a_{t^\prime}\in\{0,1\}^{\log P+O_{q,t^\prime}(1)}$ (where $2\le t^\prime\le t$), $0\le b<N$ and $0\le c\le t$, define the code
    \begin{equation*}
        \cC_{\le t}=\mathset{\bfx\in\Sigma_q^n:
        \begin{array}{c}
            \psi(\bfx)\text{ is }\text{strong-}(\ell,\epsilon)\text{-locally-balanced},   \\
             f_{P,t^\prime,0}\parenv{\bfx}=a_{t^\prime}, \forall 2\le t^\prime\le t,\\
             \vt\parenv{\psi(\bfx)}\equiv b\pmod{N},\\
             \Sum{\psi(\bfx)}\equiv c\cdot q\pmod{(t+1)q}
        \end{array}}.
    \end{equation*}
    Then $\cC_{\le t}$ is a $(\le t)$-burst-deletion correcting code. If $N=(nq+q-1)t$, there are some $a_{t^\prime}$ ($2\le t^\prime\le t$), $b$ and $c$ such that $\rho\parenv{\cC_{\le t}}\le\log n+(t-1)\log\log n+O_{q,t}(1)$.
\end{theorem}
\begin{IEEEproof}
    Suppose that $\bfx^\prime$ is obtained from a codeword $\bfx$ by a burst of at most $t$ deletions. If $\abs{\bfx^\prime}=n-1$, then only one deletion occurred. Since $N\ge(n+1)q$, one can recover $\bfx$ from $\bfx^\prime$ using the algorithm given in the proof of \Cref{lem_singledeletionpsi}.

    Now suppose $\abs{\bfx^\prime}=n-t^\prime$, where $2\le t^\prime\le t$. Denote $\bby=\psi(\bfx)$ and $\bby^\prime=\psi(\bfx^\prime)$. Suppose that the substring $x_{\sparenv{i,i+t^\prime-1}}$ was deleted from $\bfx$, where $1\le i\le n-t^\prime+1$. Then the $\bby^\prime$ is obtained from $\bby$ by replacing $\bby_{\sparenv{i,i+t^\prime}}$ with $\oplus_{k=i}^{i+t^\prime}y_k$. Let $\Delta^{\rm{bst}}=\vt(\bby)-\vt(\bby^\prime)$. Then
    \begin{equation}\label{eq_qburstdelta}
            \Delta^{\rm{bst}}=i\parenv{\Sum{\bby_{\sparenv{i,i+t^\prime}}}-\oplus_{k=i}^{i+t^\prime}y_k}+\sum_{k=1}^{t^\prime}ky_{i+k}+t^\prime\sum_{k=i+1}^{n+1-t^\prime}y_k^\prime.
    \end{equation}
    
    Suppose that $\Sum{\bby_{\sparenv{i,i+t^\prime}}}=sq+r$, where $0\le r<q$. Since $0\le\Sum{\bby_{\sparenv{i,i+t^\prime}}}\le (t^\prime+1)(q-1)$, we conclude that $0\le s\le t^\prime$. It follows from the definition of the operator ``$\oplus$" that $\oplus_{k=i}^{i+t^\prime}y_k=r$ and $\Sum{\bby_{\sparenv{i,i+t^\prime}}}-\oplus_{k=i}^{i+t^\prime}y_k=sq\le t^\prime q$. Now \Cref{eq_qburstdelta} implies that $0\le\Delta^{\rm{bst}}\le it^\prime q+\binom{t^\prime+1}{2}(q-1)+(n+1-t^\prime-i)t^\prime(q-1)=(n+1-(t^\prime-1)/2)t^\prime(q-1)+it^\prime<(n+1)t(q-1)+nt\le N$. Therefore, we can obtain the value of $\Delta^{\rm{bst}}$ by computing $\parenv{b-\vt(\bby^\prime)}\pmod{N}$.

    Let $\Delta_{sum}=\Sum{\bby}-\Sum{\bby^\prime}$. By the relationship between $\bby$ and $\bby^\prime$, we have $\Delta_{sum}=\Sum{\bby_{\sparenv{i,i+t^\prime}}}-\oplus_{k=i}^{i+t^\prime}y_k=sq\in\mathset{mq:0\le m \le t^\prime}$. So we can obtain the value of $\Delta_{sum}$ by computing $\parenv{cq-\Sum{\bby^\prime}}\pmod{(t+1)q}$. It follows from \Cref{eq_qburstdelta} that
    \begin{equation}\label{eq_qburstdeltai}
        \Delta^{\rm{bst}}=i\Delta_{sum}+\sigma^{(i)}+t^\prime\sum_{k=i+1}^{n+1-t^\prime}y_k^\prime,
    \end{equation}
where $\sigma^{(i)}=\sum_{k=1}^{t^\prime}ky_{i+k}\in\sparenv{0,\binom{t^\prime+1}{2}(q-1)}$.

Next, we will describe how to find a substring of $\bby^\prime$ which contains $\bby_i^\prime$ and has length at most $\ell$. Start from $y^\prime_{n+1-t^\prime}$ to find the first index $j$ such that there exists some $\sigma^{(j)}\in\sparenv{0,\binom{t^\prime+1}{2}(q-1)}$ such that
\begin{equation}\label{eq_qburstdeltaj}
  \Delta^{\rm{bst}}=j\Delta_{sum}+\sigma^{(j)}+t^\prime\sum_{k=j+1}^{n+1-t^\prime}y_k^\prime.
\end{equation}
By \Cref{eq_qburstdeltai}, such an index does exist. In addition, index $j$ can be found in $O(n)$ time.  Since we start from  $y^\prime_{n+1-t^\prime}$, it is obvious that $i\le j$. 
\begin{clmr}\label{clm_generalburst}
     It holds that $j-i<\ell$
\end{clmr}
\noindent\textit{Proof of Claim:} Let $A=t^\prime\sum_{k=i+t^\prime+1}^{j+t^\prime}y_k$. By the relationship between $\bby$ and $\bby^\prime$, it is easy to see that $y_k^\prime=y_{k+t^\prime}$ for any $k>i$. Then it follows from \Cref{eq_qburstdeltai,eq_qburstdeltaj} that
\begin{equation}\label{eq_qburstA}
A=t^\prime\sum_{k=i+1}^{j}y_k^\prime=\parenv{j-i}\Delta_{sum}+\sigma^{(j)}-\sigma^{(i)}.
\end{equation}

We prove the claim by contradiction. Suppose on the contrary that $j-i\ge\ell$. Since $\bby$ is strong-$(\ell,\epsilon)$-locally-balanced, we have
\begin{equation}\label{eq_qburstAbound}
    t^\prime\parenv{\frac{q-1}{2}-\epsilon}(j-i)\le A\le  t^\prime\parenv{\frac{q-1}{2}+\epsilon}(j-i).
\end{equation}
Recall that $\Delta_{sum}=sq$ for some $s\in\mathset{0,1,\ldots,t^\prime}$. By assumption, $\parenv{q,t,\epsilon}$ is a good triple. It follows from (\ref{eq_goodtriple1}) that $s_{t^\prime}q>\parenv{\frac{q-1}{2}+\epsilon}t^\prime$ and $\parenv{s_{t^\prime}-1}q<\parenv{\frac{q-1}{2}-\epsilon}t^\prime$.

Suppose that $s\ge s_{t^\prime}$. By the second inequality in (\ref{eq_qburstAbound}) and the fact that $\ell>M_{q,t,\epsilon}$, we conclude that $(j-i)\Delta_{sum}-A\ge\sparenv{s_{t^\prime}q-t^\prime\parenv{\frac{q-1}{2}+\epsilon}}(j-i)\ge\sparenv{s_{t^\prime}q-t^\prime\parenv{\frac{q-1}{2}+\epsilon}}\ell>\binom{t^\prime+1}{2}(q-1)\ge\sigma^{(i)}-\sigma^{(j)}$. This contradicts \Cref{eq_qburstA}.
If $0<s<s_{t^\prime}$, by the first inequality in (\ref{eq_qburstAbound}) and the fact that $\ell>M_{q,t,\epsilon}$, we have $A-(j-i)\Delta_{sum}\ge\sparenv{t^\prime\parenv{\frac{q-1}{2}-\epsilon}-(s_{t^\prime}-1)q}(j-i)\ge\sparenv{t^\prime\parenv{\frac{q-1}{2}-\epsilon}-(s_{t^\prime}-1)q}\ell>\binom{t^\prime+1}{2}(q-1)\ge\sigma^{(j)}-\sigma^{(i)}$. This contradicts \Cref{eq_qburstA}. \hfill$\square$

By \Cref{clm_generalburst}, the substring $\bby^\prime_{\sparenv{j-\ell+1,j}}$ of $\bby^\prime$ contains $y^\prime_i$. By the relationship between $\bby$ and $\bby^\prime$, the substring $\bby_{\sparenv{j-\ell+1,j+t^\prime}}$ of $\bby$ contains $\bby_{\sparenv{i,i+t^\prime}}$. Then by the relationship between $\bfx$ and $\bby$, the substring $\bfx_{\sparenv{j-\ell+1,j+t^\prime-1}}$ of $\bfx$ contains $\bfx_{\sparenv{i,i+t^\prime-1}}$. Clearly, this substring has length at most $P$. By \Cref{lem_Pboundedsingleburst}, the code $\cC_{\le t}$ is a $P$-bounded $t^\prime$-burst-deletion correcting code for any $2\le t^\prime\le t$. Therefore, $\bfx$ can be recovered from $\bfx^\prime$.
\end{IEEEproof}

\begin{remark}
The above theorem gives a binary $(\le 3)$-burst-deletion code with redundancy $\log n+2\log\log n+O(1)$.  The redundancy can be further reduced to $\log n + \log \log n + O(1)$. 
Let $b, N$ be defined in Theorem \ref{thm_generalburst}.  As mentioned in \Cref{sec_diff}, the code
\[
\left\{ \mathbf{x} \in \Sigma_2^n : \operatorname{VT}(\psi(\mathbf{x})) \equiv b \pmod{N} \right\}
\]
can correct a burst of at most two deletions.
Based on this, for binary alphabets with $t' \leq 2$, we can directly correct $t'$-burst-deletion. This allows us to remove the $P$-bound condition $f_{P,2,0}(\mathbf{x}) = a_2$ in Theorem \ref{thm_generalburst} for $2$-burst-deletion correction, and the resulting code still satisfies $(\leq 3)$-burst-deletion correction while requiring at most $\log n +\log \log n + O(1)$ bits of redundancy.
\end{remark}

\section{Localized-Deletion Correcting Codes}\label{sec_localized}
%%%%%%%%%%%%%%%%%%%%%%%%%%%%%%%%%%%%%%%%%%%%%%%%
Recall that a $t$-localized-deletion in a sequence refers to deleting at most $t$ symbols in a length-$t$ window of the sequence and the position of the window is not known a priori. In this section, we extend the construction in \Cref{thm_generalburst} to the $t$-localized-deletion model. 

For a good triple $(q,t,\epsilon)$ and $t\ge3$, denote by $s_{t^\prime}$ the unique integer in the interval $I_{t^\prime}$ (see \eqref{eq_openinterval}).
By the definition of good triples, it is clear that $M_{q,t,\epsilon}^\prime>0$. For each $2\le t^\prime\le t$, denote $h\parenv{t^\prime}=(q+1)(t^\prime)^2+\sparenv{(4q-2)t+q-1}t^\prime-2qt$ and
    $$
    f\parenv{t^\prime}=\frac{h\parenv{t^\prime}}{2s_{t^\prime}q-(q-1+2\epsilon)t^\prime},\quad g\parenv{t^\prime}=\frac{h\parenv{t^\prime}}{(q-1-2\epsilon)t^\prime-2(s_{t^\prime}-1)q}.
    $$
Then define
\begin{equation*}
    M_{q,t,\epsilon}^\prime=\max\mathset{f\parenv{t^\prime},g\parenv{t^\prime}:2\le t^\prime\le t}.
\end{equation*}
\begin{lemma}\label{lem_localM}
    Suppose that $(q,t,\epsilon)$ is a good triple and $t\ge3$. Let $t_1$ and $t_2$ be defined in \Cref{lem_goodtriple}.
    \begin{enumerate}[$(i)$]
        \item If $q\ge t$, then $M_{q,t,\epsilon}^\prime=\max\mathset{f(t_2),g(t_1)}$.
        \item If $2\le q<t<2q$, then $M_{q,t,\epsilon}^\prime=\max\mathset{f(t_2),g(t_1),f(q+1)}$.
    \end{enumerate}
\end{lemma}
\begin{IEEEproof}
According to \Cref{lem_goodtriple}, we have that $s_{t^\prime}=\ceilenv{t^\prime/2}$ when $t^\prime\le q$ and $s_{t^\prime}=\floorenv{t^\prime/2}$ when $t^\prime> q$.
Then it is straightforward to verify that: both $f\parenv{t^\prime}$ and $g\parenv{t^\prime}$ increase with $t^\prime$ when $t^\prime$ only takes even values; both $f\parenv{t^\prime}$ and $g\parenv{t^\prime}$ increase with $t^\prime$  when $t^\prime$ only takes odd values and $t^\prime<q$. Therefore, when $t\le q$, it holds that $M_{q,t,\epsilon}^\prime=\max\mathset{f\parenv{t_1},g(t_1),f\parenv{t_2},g\parenv{t_2}}=\max\mathset{f(t_2),g(t_1)}$. This proves (i).

When $t^\prime$ only takes odd values and $t^\prime>q$, $f\parenv{t^\prime}$ decreases with $t^\prime$ while $g\parenv{t^\prime}$ increases with $t^\prime$. Therefore, $M_{q,t,\epsilon}^\prime=\max\mathset{f(q),g(q-1),f(t_2),g(t_2),f(q+1),g(t_1)}=\max\mathset{f(t_2),g(t_1),f(q+1)}$.
\end{IEEEproof}

This main result in this section is presented below.
\begin{theorem}\label{thm_localized}
    Let $\parenv{q,t,\epsilon}$ be a good triple, which is specified in \Cref{lem_goodtriple}. Let $M_{q,t,\epsilon}^\prime$ be given in \Cref{lem_localM}. Suppose that integer $n$ is sufficiently large such that $\ell=\ceilenv{\frac{(q-1)^2}{\epsilon^2}\log_{\mathsf{e}}\parenv{2(n+1)\sqrt{q}}}>M_{q,t,\epsilon}^\prime$. Let $P=\ell+t-1$ and $N\ge(t(n+t)-1)q$. For any $2\le t^\prime\le t$, let $f_{P,t,t-t^\prime}$ be the function in \Cref{lem_Pboundedsingleburst}. For any $a_{t^\prime}\in\{0,1\}^{\log P+O_{q,t,t^\prime}(1)}$ (where $2\le t^\prime\le t$), $0\le b<N$ and $0\le c\le t$, define the code
    \begin{equation*}
        \cC_{ t}^{\rm{loc}}=\mathset{\bfx\in\Sigma_q^n:
        \begin{array}{c}
            \psi(\bfx)\text{ is }\text{strong-}(\ell,\epsilon)\text{-locally-balanced},   \\
             f_{P,t,t-t^\prime}\parenv{\bfx}=a_{t^\prime}, \forall 2\le t^\prime\le t,\\
             \vt\parenv{\psi(\bfx)}\equiv b\pmod{N},\\
             \Sum{\psi(\bfx)}\equiv c\cdot q\pmod{(t+1)q}
        \end{array}}.
    \end{equation*}
    Then the code $\cC_{ t}^{\rm{loc}}$ can correct a $t$-localized-deletion. If $N=(t(n+t)-1)q$, there are some $a_{t^\prime}$ ($2\le t^\prime\le t$), $b$ and $c$ such that $\rho\parenv{\cC_{ t}^{\rm{loc}}}\le\log n+(t-1)\log\log n+O_{q,t}(1)$.
\end{theorem}
\begin{IEEEproof}
Suppose that $\bfx$ is the transmitted codeword and $\bfx\in\Sigma_q^{n-t^\prime}\in\cD_t^{\rm{loc}}(\bfx)$ is the received sequence. Let $\bby=\psi(\bfx)$ and $\bby^\prime=\psi\parenv{\bfx^\prime}$. To decode $\bfx$ from $\bfx^\prime$, it suffices to decode $\bby$ from $\bby^\prime$.
Note that the value of $t^\prime$ is clear from the length of $\bfx^\prime$. When $t^\prime=1$, since $N\ge(n+1)q$, we can recover $\bby$ using the algorithm presented in the proof of \Cref{lem_singledeletionpsi}.
Now suppose that $t^\prime\ge 2$.

The remainder of this proof is similar to but more involved than the proof of \Cref{thm_generalburst}. To make it easier to follow, we break down the argument into four steps.

\vspace{5pt}
\noindent\textbf{Step 1:} deriving a formula analogous to \eqref{eq_qburstdeltai}.

According to \Cref{dfn_localized}, there are positive integers $t_1,\ldots,t_{k}$ and indices $1\le i_1<\cdots<i_{k}\le n-t_k+1$ satisfying: $t_1+\cdots+t_k=t^\prime$, $i_k-i_1\le t-t_k$ and $i_{j}-i_{j-1}>t_{j-1}$ for all $1<j\le k$, such that
\begin{equation}\label{eq_localized1}
    x_i^\prime=
    \left\{
    \begin{array}{lc}
         x_i,&\text{ if }i<i_1,\\
         x_{i+\sum_{j=1}^st_j},&\text{ if }i_s-\sum_{j=1}^{s-1}t_j\le i<i_{s+1}-\sum_{j=1}^st_j\\
         &\text{where }1\le s< k,\\
         x_{i+\sum_{j=1}^kt_j},&\text{ if }i_{k}-\sum_{j=1}^{k-1}t_j\le i\le n-t^\prime.
    \end{array}
    \right.
\end{equation}
Recall that $\alpha\oplus\beta\triangleq\parenv{\alpha+\beta}\pmod{q}$ for $\alpha,\beta\in\Sigma_q$. It follows from \Cref{obs_sumpsi} (i) and (\ref{eq_localized1}) that $\bby^\prime$ is obtained from $\bby$ by replacing the $k$ substrings $\bby_{\sparenv{i_s,i_s+t_s}}$ with the $k$ symbols $\oplus_{j=i_s}^{i_s+t_s}y_j$ ($s=1,\ldots,k$), respectively. Therefore, it holds that
\begin{equation}\label{eq_localized2}
y_i^\prime=
\left\{
    \begin{array}{lc}
         y_i,&\text{ if }i<i_1,\\
         \oplus_{j=i_s}^{i_s+t_s}y_j,&\text{ if }i=i_s-\sum_{j=1}^{s-1}t_j\\
         &\text{where }1\le s\le k,\\
         y_{i+\sum_{j=1}^st_j},&\text{ if }i_s-\sum_{j=1}^{s-1}t_j< i<i_{s+1}-\sum_{j=1}^st_j\\
         &\text{where }1\le s< k,\\
         y_{i+\sum_{j=1}^kt_j},&\text{ if }i_{k}-\sum_{j=1}^{k-1}t_j< i\le n-t^\prime+1.
    \end{array}
    \right.
\end{equation}
Let $\Delta^{\rm{loc}}=\vt(\bby)-\vt\parenv{\bby^\prime}$. By (\ref{eq_localized2}) and the fact that $\sum_{s=1}^kt_s=t^\prime$, we obtain
\begin{equation}\label{eq_localized32}
\begin{aligned}
    \Delta^{\rm{loc}}&=\sum_{s=1}^k\sparenv{\sum_{i=i_s}^{i_s+t_s}iy_i-\parenv{i_s-\sum_{j=1}^{s-1}t_j}\oplus_{j=i_s}^{i_s+t_s}y_j}+\sum_{s=1}^{k-1}\parenv{\sum_{j=1}^st_j}\Sum{\bby_{\sparenv{i_s+t_s+1,i_{s+1}-1}}}\\
    &\quad\quad\quad\quad+t^\prime\cdot\Sum{\bby^\prime_{\sparenv{i_k-\sum_{j=1}^{k-1}t_j+1,n-t^\prime+1}}}.
\end{aligned}
\end{equation}

Let $\Delta_{sum}=\Sum{\bby}-\Sum{\bby^\prime}$ and $\Delta_{sum}^{(s)}=\Sum{\bby_{\sparenv{i_s,i_s+t_s}}}-\oplus_{j=i_s}^{i_s+t_s}y_j$ for all $1\le s\le k$. Then it holds that $\Delta_{sum}=\sum_{s=1}^{k}\Delta_{sum}^{(s)}$. By definition, there exists some $0\le m_s\le t_s$, such that $\Delta_{sum}^{(s)}=m_sq$ for each $1\le s\le k$. This means that there exists some $0\le m\le\sum_{s=1}^kt_s=t^\prime\le t$ such that $\Delta_{sum}=mq$. Therefore, we can obtain the value of $\Delta_{sum}$ by computing $\parenv{cq-\Sum{\bby^\prime}}\pmod{(t+1)q}$.
By simple calculation, we obtain
\begin{equation}\label{eq_firstsummation}
\begin{aligned}
    &\sum_{s=1}^k\sparenv{\sum_{i=i_s}^{i_s+t_s}iy_i-\parenv{i_s-\sum_{j=1}^{s-1}t_j}\oplus_{j=i_s}^{i_s+t_s}y_j}\\
    =&i_1\Delta_{sum}+\sum_{s=1}^{k}\sum_{i=i_s}^{i_s+t_s}(i-i_s)y_i+\sum_{s=2}^{k}(i_s-i_1)\Delta_{sum}^{(s)}+\sum_{s=2}^{k}\parenv{\sum_{j=1}^{s-1}t_j}\oplus_{j=i_s}^{i_s+t_s}y_j\\
    =&i_1\Delta_{sum}+\sum_{s=1}^{k}\sum_{i=i_s}^{i_s+t_s}(i-i_s)y_i+q\sum_{s=2}^{k}(i_s-i_1)m_s+\sum_{s=2}^{k}\parenv{\sum_{j=1}^{s-1}t_j}\oplus_{j=i_s}^{i_s+t_s}y_j.
\end{aligned}
\end{equation}
Plugging this into \eqref{eq_localized32}, we conclude that $\Delta^{\rm{loc}}\ge0$. Combining \eqref{eq_firstsummation} with \eqref{eq_localized2} and \eqref{eq_localized32}, we obtain
\begin{align}
\Delta^{\rm{loc}}=i_1\Delta_{sum}+\sigma^{(i_1)}+t^\prime\cdot\Sum{\bby^\prime_{\sparenv{i_1+1,n-t^\prime+1}}}\label{eq_Deltai}
\end{align}
where
\begin{equation}\label{eq_msigma0}
    \sigma^{(i_1)}=\sum_{s=1}^{k}\sum_{i=i_s}^{i_s+t_s}(i-i_s)y_i+q\sum_{s=2}^{k}(i_s-i_1)m_s-\sum_{s=2}^{k}\parenv{\sum_{j=s}^{k}t_j}\oplus_{j=i_s}^{i_s+t_s}y_j-\sum_{s=1}^{k-1}\parenv{\sum_{j=s+1}^{k}t_j}\Sum{\bby_{\sparenv{i_s+t_s+1,i_{s+1}-1}}}.
\end{equation}

\Cref{eq_Deltai} is the formula that we desired.

\vspace{5pt}
\noindent\textbf{Step 2:} deriving bounds on $\sigma^{(i_1)}$ and $\Delta^{\rm{loc}}$, as we did for $\sigma^{(i)}$ and $\Delta^{\rm{bst}}$ in the proof of \Cref{thm_generalburst}.

Recall that $\Delta^{\rm{loc}}\ge0$. Next, we aim to upper bound $\sigma^{(i_1)}$ and $\Delta^{\rm{loc}}$. It is straightforward to verify that
\begin{equation}\label{eq_sigma1}
    0\le\sum_{s=1}^{k}\sum_{i=i_s}^{i_s+t_s}(i-i_s)y_i\le(q-1)\sum_{s=1}^{k}\sum_{i=1}^{t_s}i=(q-1)\sum_{s=1}^k\binom{t_s+1}{2}\le(q-1)\binom{t^\prime+1}{2},
\end{equation}
where the last inequality follows from the fact that $\sum_{s=1}^kt_s=t^\prime$ and $\sum_{s=1}^kt_s^2\le (t^\prime)^2$.
Since $i_s-i_1\le t-t_s$ and $m_s\le t_s$ for any $s$ and $\sum_{s=1}^kt_s=t^\prime$, we obtain
\begin{equation}\label{eq_sigma2}
    0\le\sum_{s=2}^{k}(i_s-i_1)m_s\le\sum_{s=2}^{k}(t-t_s)t_s=t(t^\prime-t_1)-\sum_{s=2}^kt_s^2\le (t^{\prime}-t_1)\parenv{t-\frac{t^{\prime}-t_1}{k-1}}\le t(t^\prime-1),
\end{equation}
where the third inequality follows from the Cauchy–Schwarz inequality. The fact that $\sum_{s=1}^kt_s=t^\prime$ also results in
\begin{equation}\label{eq_sigma3}
\begin{aligned}
0\le\sum_{s=2}^{k}\parenv{\sum_{j=s}^{k}t_j}\oplus_{j=i_s}^{i_s+t_s}y_j\le(q-1)(k-1)t^\prime
\end{aligned}
\end{equation}
and $\sum_{s=1}^{k-1}\parenv{\sum_{j=s+1}^{k}t_j}\Sum{\bby_{\sparenv{i_s+t_s+1,i_{s+1}-1}}}\le (q-1)t^\prime\sum_{s=1}^{k-1}(i_{s+1}-i_s-t_s-1)$. Since $i_k-i_1\le t-t_k$, we conclude that
$\sum_{s=1}^{k-1}(i_{s+1}-i_s-t_s-1)=i_k-i_1-\sum_{s=1}^{k-1}t_s-(k-1)\le t-t^\prime-k+1$ and thus,
\begin{equation}\label{eq_sigma4}
    0\le\sum_{s=1}^{k-1}\parenv{\sum_{j=s+1}^{k}t_j}\Sum{\bby_{\sparenv{i_s+t_s+1,i_{s+1}-1}}}\le(q-1)t^\prime(t-t^\prime-k+1).
\end{equation}
Then it follows from \eqref{eq_msigma0}--\eqref{eq_sigma4} that
\begin{equation}\label{eq_sigma5}
    -(q-1)(t-t^\prime)t^\prime\le\sigma^{(i_1)}\le\binom{t^\prime+1}{2}(q-1)+qt(t^\prime-1).
\end{equation}

Now we upper bound $\Delta^{\rm{loc}}$. By \eqref{eq_localized2} we have
\begin{equation}\label{eq_split1}
    \Sum{\bby^\prime_{\sparenv{i_1+1,n-t^\prime+1}}}=\sum_{s=1}^{k-1}\Sum{\bby_{\sparenv{i_s+t_s+1,i_{s+1}-1}}}+\Sum{\bby_{\sparenv{i_k+t_k+1,n+1}}}+\sum_{s=2}^k\oplus_{j=i_s}^{i_s+t_s}y_j.
\end{equation}
Since $\oplus_{j=i_s}^{i_s+t_s}y_j\le q-1$ for each $s$ and
$$
\sum_{s=1}^{k-1}\Sum{\bby_{\sparenv{i_s+t_s+1,i_{s+1}-1}}}+\Sum{\bby_{\sparenv{i_k+t_k+1,n+1}}}\le\parenv{n+1-t^\prime-i_1-k+1}(q-1),
$$
it follows from \eqref{eq_Deltai}, \eqref{eq_sigma5} and \eqref{eq_split1} that
\begin{equation}\label{eq_comDelta}
\begin{aligned}
    \Delta^{\rm{loc}}&\le i_1qt^\prime+(q-1)\binom{t^\prime+1}{2}+qt(t^\prime-1)+(n+1-t^\prime-i_1)(q-1)t^\prime\\
    &=\sparenv{(n+t)q-(t/t^\prime-1)q-(n+1-i_1)-\frac{t^\prime-1}{2}(q-1)}t^\prime\\
    &\le\parenv{t(n+t)-1}q-1<N.
\end{aligned}
\end{equation}
Therefore, we can obtain the value of $\Delta^{\rm{loc}}$ by computing $\parenv{b-\vt\parenv{\bby^\prime}}\pmod{N}$.

\vspace{5pt}
\noindent\textbf{Step 3:} approximately locating errors.

Recall that in \eqref{eq_Deltai}, values of $\Delta^{\rm{loc}}$ and $\Delta_{sum}$ are known to us. Start from $y^\prime_{n+1-t^\prime}$ to find the first index $j$ such that there is some $\sigma^{(j)}\in\sparenv{-(q-1)(t-t^\prime)t^\prime,\binom{t^\prime+1}{2}(q-1)+qt(t^\prime-1)}$ such that
\begin{equation}\label{eq_Deltaj}
  \Delta^{\rm{loc}}=j\Delta_{sum}+\sigma^{(j)}+t^\prime\Sum{\bby^\prime_{\sparenv{j+1,n-t^\prime+1}}}.
\end{equation}
By \eqref{eq_Deltai} and \eqref{eq_sigma5}, such an index does exist. In addition, index $j$ can be found in $O(n)$ time.  Since we start from  $y^\prime_{n+1-t^\prime}$, it is obvious that $i_1\le j$. Similar to the proof of \Cref{thm_generalburst}, we can show the following claim.
\begin{clmr}\label{clm_localized}
   It holds that $j-i_1<\ell-t^\prime+t_1$ .
\end{clmr}
\noindent\textit{Proof of Claim:} It follows from \Cref{eq_Deltai,eq_Deltaj} that
\begin{equation}\label{eq_Deltadifference}
    t^\prime\Sum{\bby^\prime_{\sparenv{i_1+1,j}}}=(j-i_1)\Delta_{sum}+\sigma^{(j)}-\sigma^{(i_1)}.
\end{equation}
Suppose on the contrary that $j-i_1\ge\ell-t^\prime+t_1$. Since $\ell>M_{q,t,\epsilon}^\prime\ge t-2$ and $i_k-i_1\le t-t_k$, we conclude that $j\ge t-1-t^\prime+t_1+i_1\ge i_k-(t^\prime-t_k)=i_k-\sum_{s=1}^{k-1}t_s$. Then by \eqref{eq_localized2}, substring $\bby^\prime_{\sparenv{i_1+1,j}}$ contains $\oplus_{r=i_s}^{i_s+t_s}y_r$, for all $2\le s\le k$. Adding $t^\prime\sum_{s=2}^{k}\parenv{\Sum{\bby_{\sparenv{i_s,i_s+t_s}}}-\oplus_{r=i_s}^{i_s+t_s}y_r}=t^\prime\sum_{s=2}^{k}\Delta_{sum}^{(s)}$ at both sides of \eqref{eq_Deltadifference} and noticing that $y_i^\prime=y_{i+\sum_{r=1}^st_r}$ for all $i_s-\sum_{r=1}^{s-1}t_r< i<i_{s+1}-\sum_{j=r}^st_r$ and $2\le s< k$, and that $y^\prime_i=y_{i+\sum_{j=1}^kt_j}$ for all $i_{k}-\sum_{j=r}^{k-1}t_r< i\le n-t^\prime+1$, we obtain
\begin{equation}\label{eq_difference}
    t^\prime\Sum{\bby_{\sparenv{i_1+t_1+1,j+t^\prime}}}=(j-i_1+t^\prime-t_1)\Delta_{sum}+\delta^{(j)},
\end{equation}
where $\delta^{(j)}=\sigma^{(j)}-\sigma^{(i_1)}+t_1\Delta_{sum}-t^\prime\Delta_{sum}^{(1)}$. Since $\sigma^{(j)},\sigma^{(i_1)}\in\sparenv{-(q-1)(t-t^\prime)t^\prime,\binom{t^\prime+1}{2}(q-1)+qt(t^\prime-1)}$, $t_1\le t^\prime$ and $\Delta_{sum},\Delta_{sum}^{(1)}\in\sparenv{0,t^\prime q}$, we have $\abs{\delta^{(j)}}\le \parenv{t^\prime}^2+\frac{(2t+t^\prime+1)t^\prime}{2}(q-1)+qt(t^\prime-1)$.

Recall that it is assumed that $j-i_1\ge\ell-t^\prime+t_1$. In this case, we have $\abs{\bby_{\sparenv{i+t_1+1,j+t^\prime}}}=j+t^\prime-i_1-t_1\ge\ell$. Since $\bby$ is strong-$(\ell,\epsilon)$-locally-balanced, it follows that
$\parenv{\frac{q-1}{2}-\epsilon}(j-i_1+t^\prime-t_1)\le\Sum{\bby_{\sparenv{i_1+t_1+1,j+t^\prime}}}\le\parenv{\frac{q-1}{2}+\epsilon}(j-i_1+t^\prime-t_1)$. Now similar to the proof of \Cref{clm_generalburst}, we can show that \Cref{eq_difference} is impossible under the assumption that $(q,t,\epsilon)$ is a good triple and $\ell>M_{q,t,\epsilon}^{\prime}$. This completes the proof of \Cref{clm_localized}.
\hfill$\square$

By \Cref{clm_localized}, the substring $\bby^\prime_{\sparenv{j-\ell+1,j}}$ of $\bby^\prime$ contains $y^\prime_{i_s-\sum_{r=1}^{s-1}t_r}$ for all $1\le s\le k$. By the relationship between $\bby$ and $\bby^\prime$, the substring $\bby_{\sparenv{j-\ell+1,j+t^\prime}}$ of $\bby$ contains $\bby_{\sparenv{i_1,i_k+t_k}}$. Then by the relationship between $\bfx$ and $\bby$, the substring $\bfx_{\sparenv{j-\ell+1,j+t^\prime-1}}$ of $\bfx$ contains $\bfx_{\sparenv{i_1,i_k+t_k-1}}$. Clearly, this substring has length at most $P$.

\vspace{5pt}
\noindent\textbf{Step 4:} correcting errors.

Since $\bfx^\prime$ results from $\bfx$ by a $t$-localized-deletion and there are exactly $t^\prime$ deletions occurred, $\bfx^\prime$ is also obtained from $\bfx$ by a $(t,t-t^\prime)$-burst-error.
By \Cref{lem_Pboundedsingleburst},  the code $\cC_{ t}^{\rm{loc}}$ is a $P$-bounded $(t,t-t^\prime)$-burst-error correcting code for any $2\le t^\prime\le t$. Therefore, $\bfx$ can be recovered from $\bfx^\prime$.
\end{IEEEproof}

\section{Encoding into Strong-$(\ell,\epsilon)$-Locally-Balanced Differential Sequences}\label{sec_algorithms}
In this section, for given $q\ge2$ and $0<\epsilon<(q-1)/2$, define
$p_1(\epsilon)\triangleq\frac{q-1}{2}-\epsilon$ and
$p_2(\epsilon)\triangleq\frac{q-1}{2}+\epsilon$.
Recall that a sequence is strong $(\ell,\epsilon)$-locally-balanced if any substring of length  $\ell^\prime\ge\ell$ has $L_1$-weight at least $p_1(\epsilon)\ell^\prime$ and at most $p_2(\epsilon)\ell^\prime$. This strong local-balance property of differential sequences serves as a foundational component in constructions given in \Cref{sec_burst,sec_localized}. This motivates our core objective in this section: developing an encoding scheme that maps an arbitrarily given sequence into another sequence whose differential sequence is strong-$(\ell,\epsilon)$-locally-balanced.

 Let $\cS_{\rm{bal}}^{(q)}\parenv{n,\ell,\epsilon}\subseteq\Sigma_q^n$ denote the set of all length-$n$ strong-$(\ell,\epsilon)$-locally-balanced sequences. It follows from \Cref{lem_slocalbalanced} that as long as $\parenv{(q-1)^2/\epsilon^2}\log_{\e}\parenv{2(n+1)\sqrt{q}}\le\ell\le n$, there are at least $q^n/2$ sequences in $\Sigma_q^n$ whose differential sequences are in $\cS_{\rm{bal}}^{(q)}\parenv{n+1,\ell,\epsilon}$. This suggests a potential encoder with only \emph{one} redundant symbol. However, we do not know how to design such an encoder. Instead, we design an encoding algorithm which converts a sequence to a sequence whose differential sequence is strong-$(\ell,\epsilon)$-locally-balanced, with \emph{two} redundant symbols. Our two-stage encoding framework operates as follows:
\begin{enumerate}[\textbf{Stage} 1]
    \item For an input $\bfx^\prime\in\Sigma_q^{n-2}$, encode its differential sequence $\psi\parenv{\bfx^\prime}\in\Sigma_q^{n-1}$ into an intermediate sequence $\bby^\prime=\cS_{\rm{bal}}^{(q)}(n,\ell-1,\eta_2)$. The constraint on parameter $\eta_2$ will be specified at the end of \Cref{subsec_step1}.
    \item Recall that a $q$-ary sequence is a differential sequence of some sequence if and only if its $L_1$ weight is congruent to $0$ modulo $q$. We further encode $\bby^\prime$ into a sequence $\bby$ such that $\bby\in\cS_{\rm{bal}}(n+1,\ell,\epsilon)$ and $\Sum{\bby}\equiv 0\pmod{q}$. Then output $\bfx=\psi^{-1}(\bby)\in\Sigma_q^n$.
\end{enumerate}
Technical implementations of these stages are detailed in \Cref{subsec_step1,subsec_step2}.

\subsection{Realization of Stage 1}\label{subsec_step1}
%%%%%%%%%%%%%%%%%%%%%%%%%%%%%%%%%
For $n\ge\ell$ and $0\le a<b\le\ell$, let $\cW^{(q)}\parenv{n,\ell,[a,b]}$ denote the set of all length-$n$ $q$-ary sequences with the property that every substring of length \emph{exactly} $\ell$ has $L_1$-weight at least $a$ and at most $b$. When $q=2$, $a\le p_1\ell$ and $b\ge p_2\ell$ for some constants $0\le p_1<1/2<p_2\le 1$, an efficient encoder/decoder pair for $\cW^{(2)}\parenv{n,\ell,[a,b]}$ was developed in \cite{Tuan2021IT}. 

Sequences in $\cW^{(q)}\parenv{n,\ell,\sparenv{p_1(\epsilon),p_2(\epsilon)}}$ are called $(\ell,\epsilon)$-\emph{locally-balanced} sequences. It is clear that strong-$(\ell,\epsilon)$-local-balance necessarily implies $(\ell,\epsilon)$-local-balance. The converse, however, does not hold, as demonstrated by the following counterexample.
\begin{example}
    Let $q=2$, $n=6$, $\ell=4$, $\epsilon=1/4$ and $\bfx=110111$. It is easy to verify that $\bfx\in\cW\parenv{n,\ell,\sparenv{0.25\ell,0.75\ell}}$. Since $\Sum{\bfx}=5\notin\sparenv{1.5,4.5}$, $\bfx$ is not strong-$(\ell,\epsilon)$-locally-balanced. 
\end{example}

Therefore, the result in \cite{Tuan2021IT} can not be directly applied to resolve our current encoding challenge. Fortunately, we have the following lemma, which generalizes \cite[Lemma 2]{Yubo2024IT}.
\begin{lemma}\label{lem_bridge}
    Given parameters $0<\eta_1,\eta_2<\frac{q-1}{2}$ satisfying $\eta_1-\frac{\eta_1^2}{(q-1)s}+\frac{q-1}{4s}\le\eta_2<\frac{q-1}{2}$ for some $s\ge 1$. If $\bfx\in\cW^{(q)}\parenv{n,\ell_1,\sparenv{p_1(\eta_1),p_2(\eta_1)}}$, then $\bfx\in\cS_{\rm{bal}}^{(q)}\parenv{n,\ell,\eta_2}$ for any $\ell\ge s\ell_1$.
\end{lemma}
\begin{IEEEproof}
    Let $\bbu$ be an arbitrary substring of $\bfx$ of length $\ell^\prime\ge\ell$. We aim to prove that $\Sum{\bbu}\in\sparenv{p_1(\eta_2)\ell^\prime,p_2(\eta_2)\ell^\prime}$.
    
    Write $\bbu$ as the concatenation of $k:=\ceilenv{\ell^\prime/\ell_1}$ substrings: $\bbu=\bbu^{(1)}\cdots\bbu^{(k-1)}\bbu^{(k)}$, where $\abs{\bbu^{(k)}}=t:=\ell^\prime-\floorenv{\ell^\prime/\ell_1}\ell_1$ and $\abs{\bbu^{(i)}}=\ell_1$ for $1\le i<k$. Then we have $\parenv{\frac{q-1}{2}-\eta_1}\ell_1\le\Sum{\abs{\bbu^{(i)}}}\le\parenv{\frac{q-1}{2}+\eta_1}\ell_1$ for $1\le i<k$, since $\bfx\in\cW^{(q)}\parenv{n,\ell_1,\sparenv{p_1(\eta_1),p_2(\eta_1)}}$. Notice that the length of $\bbu^{(k-1)}_{[t+1,\ell_1]}\bbu^{(k)}$ is exactly $\ell_1$ and $0\le\Sum{\bbu^{(k-1)}_{[t+1,\ell_1]}}\le(\ell_1-t)(q-1)$. Thus, it holds that $\max\mathset{\parenv{\frac{q-1}{2}-\eta_1}\ell_1-(\ell_1-t)(q-1),0}\le\Sum{\bbu^{(k)}}\le\min\mathset{t(q-1),\parenv{\frac{q-1}{2}+\eta_1}\ell_1}$. Then it is easy to verify that
    \begin{equation}\label{eq_lower1}
    \begin{aligned}
        \Sum{\bbu}&=\sum_{i=1}^{k}\Sum{\bbu^{(i)}}\\
        &\ge(k-1)\parenv{\frac{q-1}{2}-\eta_1}\ell_1+\max\mathset{\parenv{\frac{q-1}{2}-\eta_1}\ell_1-(\ell_1-t)(q-1),0}\\
        &=\parenv{\frac{q-1}{2}-\eta_1}(\ell^\prime-t)+\max\mathset{t(q-1)-\parenv{\frac{q-1}{2}+\eta_1}\ell_1,0}.
    \end{aligned}
    \end{equation}
    
If $t(q-1)\le\parenv{\frac{q-1}{2}+\eta_1}\ell_1$, then $t\le\parenv{\frac{1}{2}+\frac{\eta_1}{q-1}}\ell_1$. In this case, it follows from (\ref{eq_lower1}) that
\begin{align*}
    \Sum{\bbu}&\ge\parenv{\frac{q-1}{2}-\eta_1}(\ell^\prime-t)\\
    &\ge\parenv{\frac{q-1}{2}-\eta_1}\parenv{\ell^\prime-\parenv{\frac{1}{2}+\frac{\eta_1}{q-1}}\ell_1}\\
    &\overset{(a)}{\ge}\parenv{\frac{q-1}{2}-\eta_1}\parenv{\ell^\prime-\parenv{\frac{1}{2}+\frac{\eta_1}{q-1}}\frac{\ell^\prime}{s}}\\
    &=\sparenv{\frac{q-1}{2}-\parenv{\eta_1-\frac{\eta_1^2}{(q-1)s}+\frac{q-1}{4s}}}\ell^\prime\\
    &\overset{(b)}{\ge}p_1(\eta_2)\ell^\prime,
\end{align*}
where (a) follows from the assumption that $\ell^\prime\ge\ell\ge s\ell_1$, and (b) follows from the assumption that $\eta_1-\frac{\eta_1^2}{(q-1)s}+\frac{q-1}{4s}\le\eta_2$.
If $t(q-1)\ge\parenv{\frac{q-1}{2}+\eta_1}\ell_1$, $t\ge\parenv{\frac{1}{2}+\frac{\eta_1}{q-1}}\ell_1$. In this case,  it follows from (\ref{eq_lower1}) that
\begin{align*}
    \Sum{\bbu}&\ge\parenv{\frac{q-1}{2}-\eta_1}(\ell^\prime-t)+t(q-1)-\parenv{\frac{q-1}{2}+\eta_1}\ell_1\\
    &=\parenv{\frac{q-1}{2}-\eta_1}\ell^\prime+\parenv{\frac{q-1}{2}+\eta_1}\parenv{t-\ell_1}\\
    &\ge\parenv{\frac{q-1}{2}-\eta_1}\ell^\prime+\parenv{\frac{q-1}{2}+\eta_1}\sparenv{\parenv{\frac{1}{2}+\frac{\eta_1}{q-1}}\ell_1-\ell_1}\\
    &=\parenv{\frac{q-1}{2}-\eta_1}\ell^\prime-\parenv{\frac{q-1}{2}+\eta_1}\parenv{\frac{1}{2}-\frac{\eta_1}{q-1}}\ell_1\\
    &=\parenv{\frac{q-1}{2}-\eta_1}\sparenv{\ell^\prime-\parenv{\frac{1}{2}+\frac{\eta_1}{q-1}}\ell_1}\\
    &\ge p_1(\eta_2)\ell^\prime.
\end{align*}

Up to now, we have proved the lower bound $\Sum{\bbu}\ge p_1(\eta_2)\ell^\prime$. It remains to prove the upper bound $\Sum{\bbu}\le p_2(\eta_2)\ell^\prime$. Firstly, we have
\begin{equation}\label{eq_upper1}
    \begin{aligned}
        \Sum{\bbu}&=\sum_{i=1}^{k}\Sum{\bbu^{(i)}}\\
        &\le(k-1)\parenv{\frac{q-1}{2}+\eta_1}\ell_1+\min\mathset{t(q-1),\parenv{\frac{q-1}{2}+\eta_1}\ell_1}\\
        &=\parenv{\frac{q-1}{2}+\eta_1}(\ell^\prime-t)+\min\mathset{t(q-1),\parenv{\frac{q-1}{2}+\eta_1}\ell_1}.
    \end{aligned}
\end{equation}
If $t(q-1)\ge \parenv{\frac{q-1}{2}+\eta_1}\ell_1$, it follows from (\ref{eq_upper1}) that 
\begin{align*}
    \Sum{\bbu}&\le \parenv{\frac{q-1}{2}+\eta_1}\sparenv{\ell^\prime-\parenv{\frac{1}{2}+\frac{\eta_1}{q-1}}\ell_1}+\parenv{\frac{q-1}{2}+\eta_1}\ell_1\\
    &=\parenv{\frac{q-1}{2}+\eta_1}\sparenv{\ell^\prime+\parenv{\frac{1}{2}-\frac{\eta_1}{q-1}}\ell_1}\\
    &\le\parenv{\frac{q-1}{2}+\eta_1}\sparenv{\ell^\prime+\parenv{\frac{1}{2}-\frac{\eta_1}{q-1}}\frac{\ell^\prime}{s}}\\
    &=\parenv{\frac{q-1}{2}+\eta_1-\frac{\eta_1^2}{(q-1)s}+\frac{q-1}{4s}}\ell^\prime\\
    &\le p_2(\eta_2)\ell^\prime.
\end{align*}
If $t(q-1)\le \parenv{\frac{q-1}{2}+\eta_1}\ell_1$, it follows from (\ref{eq_upper1}) that 
\begin{align*}
    \Sum{\bbu}&\le \parenv{\frac{q-1}{2}+\eta_1}\parenv{\ell^\prime-t}+t(q-1)\\
    &=\parenv{\frac{q-1}{2}+\eta_1}\ell^\prime+\parenv{\frac{q-1}{2}-\eta_1}t\\
    &\le\parenv{\frac{q-1}{2}+\eta_1}\ell^\prime+\parenv{\frac{q-1}{2}-\eta_1}\parenv{\frac{1}{2}+\frac{\eta_1}{q-1}}\ell_1\\
    &=\parenv{\frac{q-1}{2}+\eta_1}\sparenv{\ell^\prime+\parenv{\frac{1}{2}-\frac{\eta_1}{q-1}}\ell_1}\\
    &\le p_2(\eta_2)\ell^\prime.
\end{align*}
Now the proof is completed.
\end{IEEEproof}

\Cref{lem_bridge} bridges local-balance property and strong local-balance property. Therefore, to implement Stage 1, it is sufficient to encode $\psi\parenv{\bfx^\prime}$ into $\cW^{(q)}\parenv{n,\frac{\ell-1}{s},\sparenv{p_1(\eta_1),p_2(\eta_1)}}$, where $s$ is given in \Cref{lem_bridge}. As mentioned before, an efficient encoder for $\cW^{(2)}\parenv{n,\ell,[a,b]}$ was presented in \cite[Section III]{Tuan2021IT}. This algorithm is based on Corollary 1, Theorems 5 and 6 in \cite{Tuan2021IT}. Although those results were proved for binary alphabet, they can be directly generalized to $q$-ary alphabets. We respectively present these generalizations below in \Cref{prop_numSW,prop_forbidden,prop_lastwindow} and omit their proofs. In the following, let $0<\epsilon_0<(q-1)/2$.
\begin{proposition}\label{prop_numSW}
    When $m\ge\max\mathset{2q/(q-1),((q-1)^2/\epsilon_0^2)\log_{\mathsf{e}}m}$, it holds that
    $\abs{\cW^{(q)}\parenv{m,m,\sparenv{p_1(\epsilon_0)m,p_2(\epsilon_0)m}}}\ge q^{m-1}$.
\end{proposition}

Let $\cF^{(q)}\parenv{m,\epsilon_0}\subseteq\Sigma_q^m$ denote the set of sequences whose $L_1$-weight does not belong to the interval $\sparenv{p_1(\epsilon_0)m,p_2(\epsilon_0)m}$. We call $\cF^{(q)}\parenv{m,\epsilon_0}$ the set of \emph{forbidden} sequences of length $m$.
\begin{proposition}\label{prop_forbidden}
    Suppose that $n\ge 2q^3$ and $((q-1)^2/\epsilon_0^2)\log_{\e} n\le m\le C\log_{q} n$ for some constant $C>0$. Let $k=m-3-\log_q n$. There exists an injective mapping $\Phi_{m}:\cF^{(q)}\parenv{m,\epsilon_0}\rightarrow\Sigma_q^k$.
\end{proposition}

Let $\cG^{(q)}\parenv{m+1,\epsilon_0}$ denote the set of $q$-ary sequences of length $m+1$ that contains at least one forbidden substring in $\cF^{(q)}\parenv{m,\epsilon_0}$.
\begin{proposition}\label{prop_lastwindow}
    Suppose that $m\ge\max\mathset{2q^2-1,((q-1)^2/\epsilon_0^2)\log_{\mathsf{e}}(m-2)+2}$ and $m\le C\log_{q} n$ for some constant $C>0$. Then we have $\abs{\cG^{(q)}\parenv{m+1,\epsilon_0}}\le q^{m-3}$. In addition, there exists an injective mapping
    $$
    \Psi:\cG^{(q)}\parenv{m+1,\epsilon_0}\rightarrow\cW^{(q)}\parenv{m-2,m-2,\sparenv{p_1(\epsilon_0)(m-2),p_2(\epsilon_0)(m-2)}}.
    $$
\end{proposition}

With \Cref{prop_numSW,prop_forbidden,prop_lastwindow} in hand, it is easy to see that the encoding/decoding algorithms in \cite[Section III-C]{Tuan2021IT} also work for general alphabets. Now we can implement Stage 1.

\vspace{3pt}
\subsubsection{\textbf{Implementing Stage 1}}
Recall that the input is $\bfx^\prime\in\Sigma_q^{n-2}$.
Choose two real numbers $0<\eta_1,\eta_2<(q-1)/2$ and integer $s\ge 1$, such that $\eta_1-\frac{\eta_1^2}{(q-1)s}+\frac{q-1}{4s}\le\eta_2<\epsilon<\frac{q-1}{2}$. Suppose that $n\ge2q^3$ and integer $\ell$ satisfies (the exact value of $\ell$ will be specified at the end of \Cref{subsec_step2})
$$
\max\mathset{2q^2-1,\frac{(q-1)^2}{\eta_1^2}\log_{\e}n,}\le\frac{\ell-1}{s}\le C\log_q n
$$
for some constant $C>0$. Then we can apply the algorithm in \cite{Tuan2021IT} to encode $\psi\parenv{\bfx^\prime}$ into a sequence $\bby^\prime$. By \Cref{lem_bridge}, $\bby^\prime$ is also strong-$(\ell-1,\eta_2)$-locally-balanced.

Before proceeding to next subsection, we explain the lower bound on $\ell$. By \Cref{lem_goodtriple}, to encode $\psi\parenv{\bfx^\prime}$ into a strong-$(\ell-1,\eta_2)$-locally-balanced sequence, it suffices to encode $\psi\parenv{\bfx^\prime}$ into a sequence in $\cW^{(q)}\parenv{n,\frac{\ell-1}{s},\sparenv{p_1(\eta_1),p_2(\eta_1)}}$. Let $\epsilon_0=\eta_1$. Setting $m=(\ell-1)/s$ in \Cref{prop_forbidden,prop_lastwindow}, we obtain the lower bound claimed above. The bound on $\ell$ is critical in applying the encoder for $\cW^{(q)}\parenv{n,\frac{\ell-1}{s},\sparenv{p_1(\eta_1),p_2(\eta_1)}}$ given in \cite{Tuan2021IT}.

\subsection{Realization of Stage 2}\label{subsec_step2}
%%%%%%%%%%%%%%%%%%%%%%%%%%%%%%%%%%%%%%
\begin{lemma}\label{lem_SBextension}
    Let $0<\epsilon_1<\epsilon_2<(q-1)/2$. Let $m$ be sufficiently large such that $(\epsilon_2-\epsilon_1)m\ge(q-1)/2-\epsilon_1$. Suppose that $\bbu$ is strong-$(m-1,\epsilon_1)$-locally-balanced. Then for any $a\in\Sigma_q$, sequence $\bbu a$ (or $a\bbu$) is strong-$(m,\epsilon_2)$-locally-balanced.
\end{lemma}
\begin{IEEEproof}
We only prove the conclusion for sequence $\bbu a$. Similar argument can be applied to $a\bbu$.
Let $\bbv$ be a substring of $\bbu a$ of length $m^\prime\ge m$. We need to show that $\Sum{\bbv}\in\sparenv{p_1(\epsilon_2)m^\prime,p_2(\epsilon_2)m^\prime}$.

Firstly, suppose that $\bbv$ is a substring of $\bbu$. Then by assumption, we have $\Sum{\bbv}\in\sparenv{p_1(\epsilon_1)m^\prime,p_2(\epsilon_1)m^\prime}$. Since $\epsilon_1<\epsilon_2$, it follows that $p_1(\epsilon_2)m^\prime\le p_1(\epsilon_1)m^\prime\le\Sum{\bbv}\le p_2(\epsilon_1)m^\prime\le p_2(\epsilon_2)m^\prime$.

Now suppose that $\bbv=\bbu^{\prime}a$, where $\bbu^{\prime}$ is a length-$(m^\prime-1)$ suffix of $\bbu$. Since $m^\prime-1\ge m-1$, we have
\begin{equation}\label{eq_strong1}
    \parenv{\frac{q-1}{2}-\epsilon_1}\parenv{m^\prime-1}\le\Sum{\bbu^\prime}\le \parenv{\frac{q-1}{2}+\epsilon_1}\parenv{m^\prime-1}.
\end{equation}
By assumption, we have $(\epsilon_2-\epsilon_1)m^\prime\ge(\epsilon_2-\epsilon_1)m\ge(q-1)/2-\epsilon_1$ and thus, $(\epsilon_2-\epsilon_1)m^\prime+(q-1)/2+\epsilon_1\ge q-1$. Then it follows from (\ref{eq_strong1}) that
\begin{align*}
    \Sum{\bbv}&=\Sum{\bbu^\prime}+a\\
    &\ge\parenv{\frac{q-1}{2}-\epsilon_1}\parenv{m^\prime-1}+a\\
    &\ge\parenv{\frac{q-1}{2}-\epsilon_1}\parenv{m^\prime-1}-\sparenv{(\epsilon_2-\epsilon_1)m^\prime-(q-1)/2+\epsilon_1}\\
    &=p_1(\epsilon_2)m^\prime
\end{align*}
and
\begin{align*}
    \Sum{\bbv}&=\Sum{\bbu^\prime}+a\\
    &\le\parenv{\frac{q-1}{2}+\epsilon_1}\parenv{m^\prime-1}+a\\
    &\le\parenv{\frac{q-1}{2}+\epsilon_1}\parenv{m^\prime-1}+\frac{q-1}{2}+\epsilon_1+(\epsilon_2-\epsilon_1)m^\prime\\
    &=p_2(\epsilon_2)m^\prime.
\end{align*}
Now the proof is completed.
\end{IEEEproof}

\vspace{3pt}
\subsubsection{\textbf{Implementing Stage 2}} At the end of \Cref{subsec_step1}, we have encoded $\psi\parenv{\bfx^\prime}$ into sequence $\bby^\prime$, which is strong-$(\ell-1,\eta_2)$-locally-balanced. Let $\epsilon_1=\eta_2$ and $\epsilon_2=\epsilon$ in \Cref{lem_SBextension}. Suppose that $\ell$ is sufficiently large such that $(\epsilon-\eta_2)\ell\ge(q-1)/2-\eta_2$. Let $a=\parenv{q-\Sum{\bby^\prime}}\pmod{q}$. Encode $\bby^\prime$ into $\bby=\bby^\prime a$. Clearly, it holds that $\Sum{\bby}\equiv 0\pmod{q}$. In addition, we have $\bby\in\cS_{\rm{bal}}(n+1,\ell,\epsilon)$ by \Cref{lem_SBextension}. Then our encoder outputs $\psi^{-1}(\bby)$.

It remains to determine the value of $\ell$ and the lower bound of $n$. To that end, suppose that $n\ge2q^3$ is sufficiently large such that $\frac{(q-1)^2}{\eta_1^2}\log_{\e}n\ge2q^2-1$. Then according to the anaysis at the end of \Cref{subsec_step1}, we can choose $\ell=\ceilenv{s\frac{(q-1)^2}{\eta_1^2}\log_{\e}n}+1$. By \Cref{lem_SBextension}, $n$ should be large enough such that $(\epsilon-\eta_2)\ceilenv{s\frac{(q-1)^2}{\eta_1^2}\log_{\e}n}\ge\frac{q-1}{2}-\eta_2$.

\vspace{3pt}
\subsubsection{\textbf{Time Complexity}} Let $C$ be the smallest number such that $\ell\le C\log_{q} n$. It is easy to see that the running time of the proposed encoding algorithm is dominated by the running time of Stage 1, which in turn is dominated by the while loop in the algorithm presented in \cite{Tuan2021IT}. The while loop runs $O(n)$ times. The time complexity of each iteration of the while loop is $O\parenv{n^{C-1}}$. Therefore, the overall time complexity is $O\parenv{n^{C}}$.

\vspace{5pt}
Since there is an encoder for $\cW^{(q)}\parenv{n,\frac{\ell-1}{s},\sparenv{p_1(\eta_1),p_2(\eta_1)}}$ \cite{Tuan2021IT}, the decoding from $\psi^{-1}(\bby)$ back to the input $\bfx^\prime$ is straightforward. Again, the time complexity is $O\parenv{n^{C}}$.

\section{Conclusion}\label{sec_conclusion}
%%%%%%%%%%%%%%%%%%%%%%%%%%%%%%
This paper presents novel constructions of $(\leq t)$-burst-deletion-correcting and $t$-localized-deletion-correcting codes through a new position-estimation methodology. In contrast to prior works, our position-estimation approach offers enhanced simplicity. We believe this framework may find broader applications in code construction.

Current state-of-the-art constructions for these codes rely primarily on a two-phase approach: first roughly locating errors and then correcting errors locally. This inherently yields an $O(\log\log n)$ redundancy term. It remains open whether the $\log n + \Omega(1)$ lower bound on redundancy is asymptotically tight. Resolving this fundamental question may require novel insights.

\bibliographystyle{IEEEtran}
\bibliography{ref}
\end{document}